\documentclass[pra,twocolumn,showpacs,preprintnumbers]{revtex4}
\usepackage{graphicx}
\usepackage{amsfonts}
\usepackage{amssymb}
\usepackage{amsmath}
\makeindex
\newcommand{\be}{\begin{equation}}
\newcommand{\ee}{\end{equation}}
\newcommand{\bea}{\begin{eqnarray}}
\newcommand{\eea}{\end{eqnarray}}
\newcommand{\Eq}[1]{Eq.\,(\ref{#1})}% \Eq{eqn:abc}
\newcommand{\Fig}[1]{Fig.\,\ref{#1}}% \Fig{fig:abc}
\newcommand{\Sec}[1]{Sec.\,\ref{#1}}% \Sec{sec:abc} 
\newcommand{\Onlinecite}[1]{Ref.\,\onlinecite{#1}} % \Onlinecite{abc}
\newcommand{\GFk}{\hat{\mathbf{G}}_k}
\newcommand{\GFOk}{\hat{\mathbf{G}}^{(1)}_k}
\newcommand{\GFTk}{\hat{\mathbf{G}}^{(2)}_k}

\newcommand{\GFkOD}{G_k}
\newcommand{\Qn}{\hat{\mathbf{Q}}_n}
\newcommand{\br}{\mathbf{r}}
\newcommand{\brd}{\mathbf{\dot{r}}}
\newcommand{\bk}{\mathbf{k}}

\newcommand{\bE}{\mathbf{E}}

\newcommand{\bs}{\boldsymbol{\sigma}}
\newcommand{\En}{\mathbf{E}_n}

\newcommand{\Em}{\mathbf{E}_{m}}

\newcommand{\Fn}{\mathbf{F}_n}
\newcommand{\Epn}{\mbox{\boldmath ${\cal E}$}\hspace*{-1.5pt}_\nu}
\newcommand{\Ep}{\mbox{\boldmath ${\cal E}$}}

\newcommand{\heps}{\hat{\boldsymbol{\varepsilon}}}

\newcommand{\eps}{\varepsilon}
\newcommand{\hsigma}{\hat{\boldsymbol{\sigma}}}

\newcommand{\epw}{\epsilon_{\rm \omega}}

\newcommand{\kmax}{k_{\rm max}}

\newcommand{\GFkfsOD}{G^{fs}_k}
\newcommand{\AnOD}{A_n}
\newcommand{\An}{\mathbf{A}_n}
\newcommand{\EnOD}{E_n}
\newcommand{\GFkfs}{\hat{\mathbf{G}}^{fs}_k}

\newcommand{\Qm}{\hat{\mathbf{Q}}_m}
\newcommand{\Fm}{\mathbf{F}_m}

\begin{document}
%\pagewiselinenumbers
\title{Resonant-state-expansion Born approximation for waveguides with dispersion}
\author{ M.\,B. Doost}
%\author{ W.\,Langbein}
%\author{E.\,A. Muljarov}
\affiliation{Independent Researcher}
\begin{abstract}
The resonant-state expansion (RSE) Born approximation, a rigorous perturbative method developed for electrodynamic and quantum mechanical open systems, 
is further developed to treat waveguides with a Sellmeier dispersion. For media that can be described by these types of dispersion over the relevant frequency 
range, such as optical glass, I show that the perturbed RSE problem can be solved by diagonalizing a second-order eigenvalue problem. In the case of a single resonance 
at zero frequency, this is simplified to a generalized eigenvalue problem. Results are presented using analytically solvable planar waveguides and parameters of borosilicate BK7 glass, for 
a perturbation in the waveguide width. The efficiency of using either an exact dispersion over all frequencies or an approximate dispersion over a narrow frequency range is compared. 
I included a derivation of the RSE Born approximation for waveguides to make use of the resonances calculated by the RSE, an RSE extension of the well-known Born approximation.

\end{abstract}
\date{\today}
\maketitle
\section{Introduction}
Fundamental to scattering theory, the Born approximation consists of taking the incident field in place of the total field as the driving field at each point inside the scattering potential, it was 
first discovered by Born and presented in  Ref.~\cite{Born26}. The Born approximation gave an expression for the differential scattering cross section in terms of the Fourier transform of the 
scattering potential. An important feature of this appearance of the Fourier transform is the availability of the inverse Fourier transform operation for the inverse scattering problem.
The Born approximation is only valid for weak scatterers. In this paper I apply the RSE Born approximation Ref.\cite{DoostARX15,Ge14} to optical fibers or general open waveguide systems,
which allows an arbitrary number of resonant-states (RSs) 
to be taken into account for scattering and transmission perturbative calculations.

Optical fibers provide a well-controlled optical path which can carry light over potentially very long distances of several hundred to thousands of kilometers since the light is contained within the fiber by total 
internal reflection. Fibers are clearly important for telecommunication since the light may be modulated to carry information. It has recently been reported that a hollow-core photonic-band gap fiber yielded a record 
combination of low loss and wide bandwidth \cite{Poletti13}. Beyond telecommunications waveguides are used in integrated optical circuits \cite{John09}, and terabit chip-to-chip interconnects \cite{Gonzalez12}.

Critical to understanding the response of a waveguide to being optically driven are its resonant states (RSs). The concept of RSs was first conceived and used by Gamow in 1928 in order to describe mathematically the process of radioactive decay, specifically the escape from the nuclear potential of an $\alpha$ particle 
by tunneling. Mathematically this corresponded to solving Schr\"odinger's equation for outgoing boundary conditions (BCs). These states have complex frequency $\omega$ with negative imaginary part meaning 
their time dependence $\exp(-i\omega t)$ decays exponentially, thus giving an explanation for the exponential decay law of nuclear physics. The consequence of this exponential decay with time is that the 
further from the decaying system at a given instant of time, the greater the wave amplitude. An intuitive way of understanding this divergence of wave amplitude with distance is to notice that waves that are 
further away have left the system at an earlier time when less of the particle probability density had leaked out. There already exists numerical techniques for finding eigenmodes, such as the finite element 
method (FEM) and finite difference in time domain (FDTD) method to calculate resonances in open cavities. However, determining the effect of perturbations, which break the symmetry, presents a significant 
challenge as these popular computational techniques need large computational resources to model high-quality modes. Also these methods generate spurious solutions, which would damage the accuracy of the 
RSE Born approximation if included in the perturbation expansion basis.

In order to calculate the resonances of open systems we have recently developed the resonant-state expansion. Such an approach was not previously available due to the lack of a normalization for resonant states. 
I derived the normalization of resonant states as a contribution to Ref.\cite{DoostPRA14}, and the generalization of the normalisation to dispersive media I derived straight forwardly at the time of my rigorous
derivation. The problem of normalising 
resonant states stems from the fact that RSs with complex frequencies have wave functions which are exponentially growing in space away from system, making a 
volume integral over all space as would be done for a hermitian system a meaningless exercise (see Appendix F). 
However the ${\it S}$-wave normalization was previously available and analytically correct but numerically unstable as I showed analytically in Ref \cite{DoostARX15}. 

So far the RSE has been applied to non dispersive systems of different dimensionality \cite{MuljarovEPL10,DoostPRA12,DoostPRA13,ArmitagePRA14,DoostPRA14}. However, almost all realistic systems have a relevant frequency 
dispersion of the refractive index. I have recently found \cite{Doost15,DoostARX15} that an Ohm's law dispersion, i.e. a term in the susceptibility scaling at the inverse frequency, can be introduced to the RSE while keeping its 
linearity. We found in Ref.\cite{Doost15} this dispersion can be a reasonable approximation for some materials over a limited range of wavelengths such as SHOTT BK7 glass over the optical range. In this work I 
generalize the RSE approach for waveguides detailed in Ref.\cite{ArmitagePRA14} to systems constructed from dispersive media. Specifically I treat dispersion fully described by the Sellmeier equation, using a method similar 
to Muljarov {\it et al} in Ref.\cite{Egor} for nano particles obeying Drude-Lorentz dispersion, however in this case generalised to inclined geometry. I also treat 
dispersion linear in wavelength squared, a generalisation of my work in Ref.\cite{Doost15,DoostARX15} to inclined geometry. I find the generalization of my Ohm's law approach to inclined geometry to be greatly superior to the generalization
of the full
dispersion treatment. This is most likely due to the unphysicality of the full dispersion at high and low frequencies, which are beyond the fitting range of the dispersion models. In order to make use of the RSs, 
I derive the RSE Born approximation for waveguides \cite{DoostARX15,Ge14}, a theory for finding the field outside of a waveguide being internally or externally driven (see Appendix G).

The paper is organized as follows, \Sec{sec:RSE} outlines the general recursive solution of Maxwell's equation using Green's function. 
\Sec{sec:SingleSell} develops the solution in \Sec{sec:RSE} into a perturbation theory for waveguides with linear dispersion in wavelength squared. 
\Sec{sec:GenSel} develops the solution in \Sec{sec:RSE} into a perturbation theory 
for waveguides with Sellmeier dispersion. \Sec{sec:Unperturbed} 
demonstrates and compares the two approaches for including dispersion into the RSE. The perturbation considered corresponds to a narrowing of the waveguide and is discussed in \Sec{subsec:system}. In \Sec{sec:SRZ} 
I show the results using the simple dispersive approximation BK7. In \Sec{subsec:SellDis} I use the full Sellmeier dispersion of BK7. Finally, we discuss the comparison and performance of the two methods in 
\Sec{sec:Performance}, in \Sec{RSEBORN22}, I derive the equations for the RSE Born approximation for planar waveguides. In Appendix C, D, E, and F I generalize my part of the proof of the normalization of RSs to open waveguide systems.
\section{Resonant state expansion for waveguides and non-normal incidence}\label{sec:RSE}

In this section I develop the general recursive solution to the problem of calculating the resonant states of a perturbed waveguide. The recursive solution requires the Green's function (GF) of the unperturbed 
waveguide, and a perturbation which is retaining the translational invariance along the waveguide.  

I first consider a waveguide of thickness $2a$ in vacuum with translational invariance in one direction, having the dielectric constant
\be \heps_{\omega}(\brd )=\left\{
\begin{array}{cl}
\epw & \text{for\ \  } |\brd | \leqslant a\,,\\
1 & \text{for\ \  } |\brd | > a\,,
\end{array} \right.\ee
where $\epw$ is the frequency dependent relative permittivity (RP) of the wave guide with the angular frequency $\omega$. $\brd$ are the coordinates normal to the waveguide, which for a cylindrical waveguide is given in polar coordinates 
$\brd=(\rho,\theta)$ and for a planar waveguide $\brd$ is Cartesian $x$. I assume a relative permeability of $\mu=1$ throughout this work. The electric field ${\bf \hat{E}}$ satisfies Maxwell's equation,
\be\left[-\nabla\times\nabla\times-\heps_{\omega}(\br)\frac{1}{c^2}\frac{\partial^2}{\partial t^2}\right]{\bf \hat{E}}({\bf r},t)=0\,,
\label{ME}\ee
For cylindrical and planar waveguides, due to translational invariance in the $z$ direction I first assume then prove that ${\bf \hat{E}}$ can be factorised as follows 
\be {\bf \hat{E}}({\bf r},t)=e^{i(pz-\omega t)} {\bf E}(\brd)\,,
\label{EF}\ee
in which $p$ is the wave vector along the translationally invariant direction. For the component ${\bf E}(\brd)$ of the electric field, \Eq{ME} transforms to a one-dimensional (1D) wave equation in the planar case or a 2D wave-equation 
for the cylindrical waveguide to
\be\left[-L+\heps_{\omega}(\brd)\frac{\omega^2}{c^2}\right]{\bf E}(\brd)=0\,. \label{ME1D}\ee
Here $L$ is a linear operator not dependent on $z$  or  $\heps_{\omega}(\brd)$. The form of $L$ is derived in Appendix C, the derivation shows that the trial solution \Eq{EF} is correct since it demonstrates that 
${\bf \hat{E}}$ can be factorised by separation of variables. In principle $L$ could be any linear operator independent of both $z$ and $\heps_{\omega}(\brd)$, and hence the treatment of waveguides in this paper is readily applicable to quantum mechanics 
or acoustics.

The non-normal incidence, characterized by $p \neq 0$, is treated here. The previously used spectral representation of the GF in the frequency domain contains a cut for $p\neq 0$, which can be removed by mapping the 
problem onto the complex normal wave-vector space $k$, as demonstrated in Ref.\cite{ArmitagePRA14}. The relation between $k$, $p$, and $\omega$ is defined by us to be $\omega^2/c^2=k^2+p^2$, $k$ and $p$ are orthogonal 
by Pythagorean identity. I follow also here the approach used in Ref.\cite{ArmitagePRA14} and formulate the RSE in the complex $k$-plane, for which the spectral representation of the GF of an infinite planar system 
with an in-plane momentum $p\neq0$ written in the spectral form 
\be
{\hat{\mathbf{G}}^{(1)}_k}(\brd,\brd ')=\sum_n\frac{{\bf E_n}(\brd)\otimes {\bf E_n}(\brd ')}{2k_{n}(k-k_{n})}\,,
\label{GF1}
\ee
where ${\bf E_n}(\brd)$ is the electric field of a RS, defined as an eigensolution of \Eq{ME1D} with an arbitrary profile of $\heps_{\omega}(\brd )$ within a region $|\brd|<a$, satisfying the outgoing wave 
boundary conditions at infinity. The dyadic product of the ${\bf E_n}(\brd)$ fields is denoted by $\otimes$. The in-plane eigenfrequency $k_n$ are the poles in ${\hat{\mathbf{G}}_k}(\brd,\brd ')$. By definition 
$\omega^2_n/c^2=k_n^2+p^2$ where $\omega^2_n$ is the eigenfrequency corresponding to the eigenstate ${\bf E_n}(\brd)$. For a study of the derivation of such spectral representations see 
Appendix D.

The GF satisfies the equation
\be
\left[-L+\heps_{\omega}(\brd )(k^2+p^2)\right]{\hat{\mathbf{G}}_k}(\brd,\brd ')=\hat{\mathbf{1}}\delta(\brd-\brd ') \label{GFequ} \ee
In the present work, we consider the refractive profile (RP) having the property
\be \lim_{\omega\rightarrow\infty}\heps_\omega(\brd)=\heps(\brd), \label{eqn:epslim} \ee
where $\heps(\brd)$ if the frequency independent part of $\heps_\omega(\brd)$. If we assume $\heps_\omega(\brd)$ to be discontinuous through the boundary of the system then by substituting \Eq{GF1} into \Eq{GFequ}, 
convoluting with a finite function, and letting $k\rightarrow\infty$, we obtain the  sum rule (see Appendix E)
\be
\sum_n\frac{{\bf E_n}(\brd)\otimes {\bf E_n}(\brd ')}{k_{n}}=0\,,
\label{sum_rule}
\ee
which allows us to re-write the Green's function a second way as
\be
{\hat{\mathbf{G}}^{(2)}_k}(\brd,\brd ')=\sum_n\frac{{\bf E_n}(\brd)\otimes {\bf E_n}(\brd ')}{2k_{n}}\left[\frac{1}{k-k_{n}}-\dfrac{k}{k^2+p^2}\right]\,,\label{GF2} \ee
I now consider an arbitrary perturbation $\Delta\heps_{\omega}(\brd)$ of the dielectric constant inside the layer $|\brd |<a$.
I use the \Eq{GFequ} to solve the perturbed problem,
\be \Epn(\brd)=-\frac{\omega^2}{c^2} \int \GFk(\brd,\brd ')\Delta\heps_{\omega}(\brd ')\Epn(\brd ') d{\bf \brd}'\,,
\label{GFsol2} \ee
which is the solution to the equation
\be L\Epn(\brd)=\frac{\omega^2}{c^2}\left[\heps_{\omega}(\brd )+\Delta\heps_{\omega}(\brd)\right]\Epn(\brd)
\label{GFequ2} \ee
Note that the perturbed modes $\Epn(\brd)$ satisfy \Eq{ME1D} with $\heps_\omega(\brd)$ replaced by $\heps_\omega(\brd)+\Delta\heps_\omega(\brd)$ and the outgoing boundary conditions with $\omega_n$ replaced by $\omega$.
I show the solution of this equation perturbatively for two different types of dispersion in the following two sections.

\section{Single Sellmeier resonance at zero frequency}\label{sec:SingleSell}

For a Sellmeier dispersion with a single resonance at zero frequency (SRZ),  I write the perturbation in \Eq{GFsol2} as
\be \Delta\heps_\omega(\brd)=\Delta\heps(\brd)+\frac{c^2\Delta\hsigma(\brd)}{\omega^2}\,,
\label{del-eps}
\ee
again present only inside the layer $|\brd |<a$. It is then possible to linearise the RSE by using the different forms of the Green's function, similar to Ref.\cite{Egor}, \Eq{GF1} and \Eq{GF2} for the different 
components of the perturbation 
in \Eq{GFsol2},
\bea
\Ep(\brd)&=&-\int \GFOk(\brd,\brd ')\Delta\hsigma(\brd ')\Ep(\brd ') d{\bf \brd}' \nonumber\\
&&-\frac{\omega^2}{c^2} \int \GFTk(\brd ,\brd ')\Delta\heps(\brd ')\Ep(\brd ') d{\bf\brd}'\,, \label{Dyson2_S}
\eea
which results in the following relationship between unperturbed and perturbed modes
\bea
\Ep(\brd)&=&-\left(k^2+p^2\right)\sum_n\frac{{\bf E_n}(\brd)}{2k_n}\left[\frac{1}{k-k_n}-\dfrac{k}{k^2+p^2}\right] \nonumber\\
&&\times \int^{a}_{-a}{\bf E_n}(\brd ')\Delta\heps(\brd ')\Ep(\brd ')d\brd \nonumber\\
&&-\sum_n\frac{{\bf E_n}(\brd)}{2k_n(k-k_n)} \nonumber\\
&&\times\int^{a}_{-a}{\bf E_n}(\brd ')\Delta\hsigma(\brd ')\Ep(\brd ')d\brd\,. \label{Dyson2_S1}
\eea
The perturbed mode $\Ep(\brd)$ satisfies \Eq{ME1D} with $\heps_{\omega}(\brd)$ replaced by $\heps_{\omega}(\brd)+\Delta\heps_{\omega}(\brd)$ and outgoing boundary conditions with the wave vector $k$.
In the interior region $\left| \brd\right|<a$ which contains the perturbation, the perturbed RSs $\Epn$ of wavenumbers $\varkappa_\nu$ can be expanded into the unperturbed ones, 
exploiting the completeness of the latter which follows from \Eq{eqn:SR5} (see also Appendix E),
\be \Epn(\brd)=\sum_{n}b_{n\nu}{\bf E_n}(\brd)\,. \label{superp} \ee
Substituting this expansion into \Eq{Dyson2_S} and equating coefficients at the same basis functions ${\bf E_n}$ results in the matrix equation
%
%expanding $\Ep(\brd)$ as in \Eq{superp} and equating coefficients at the same basis functions ${\bf E_n}$ yields 
%
\be b_{n\nu}=\sum_m -b_{m\nu}\frac{(p^2+\varkappa_\nu k_n)V_{nm}+S_{nm}}{2k_n(\varkappa_\nu-k_n)}\,, \label{Dyson3_S} \ee
where
\be V_{nm}=\int{{\bf E_n}(\brd)\cdot\Delta\heps(\brd){\bf E_m}(\brd)\,d\brd}
\label{Vnm} \ee
and
\be S_{nm}=\int{{\bf E_n}(\brd)\cdot\Delta\hsigma(\brd){\bf E_m}(\brd)\,d\brd}. \label{Snm} \ee
With the substitution $c_{n\nu}=b_{n\nu}\sqrt{k_n}$, \Eq{Dyson3_S} can be rewritten as
\begin{multline}
0=\sum_m c_{m\nu}\left\{\varkappa_\nu\left(\!\frac{\delta_{nm}}{k_n}\!+\!\frac{V_{nm}}{2\sqrt{k_n}\sqrt{k_m}}\!\right)\right. \\\left.+\frac{p^2 V_{nm}+S_{nm}}{2k_n\sqrt{k_n}\sqrt{k_m}}-\delta_{nm}\right\}
\label{Dyson4}
\end{multline}
care should be taken to take the sign of $\sqrt{k_a}$ consistently between matrix elements. \Eq{Dyson4} is {\it linear} in $\varkappa_\nu$ and can be solved by libraries for generalized linear matrix eigenvalue problems. 
In the absence of dispersion, $S_{nm}=0$, and \Eq{Dyson4} reverts back to the expression for non-dispersive waveguides \cite{ArmitagePRA14}. 
In the absence of $p$, $pa=0$, we see that \Eq{Dyson4} reverts to an expression for dispersive perturbation to nano-particles \cite{Doost15}.

\section{Sellmeier dispersion}\label{sec:GenSel}

In this section I develop the existing RSE for wave guides into a perturbation theory for waveguides with Sellmeier dispersion, following a similar approach to Ref.\cite{Egor}. The Sellmeier dispersion is the sum of a set 
of Lorentzians in frequency with poles $q_j$ on the real axis. The starting point for this derivation is \Eq{GFsol2}, the recursive solution for the perturbed problem derived in \Sec{sec:RSE}.

In this section, we consider the RP
\be \heps_{\omega}(\brd)+\sum_j\dfrac{c^2\hsigma_j(\brd)}{\omega^2-\Omega^2_j}, \label{eqn:eps} \ee
and perturbation,
\be \Delta\heps_\omega(\brd)=\Delta\heps(\brd)+\sum_j\dfrac{c^2\Delta\hsigma_j(\brd)}{\omega^2-\Omega^2_j}, \label{eqn:eps22} \ee
with $j$ numbering the resonances at frequencies $\Omega_j$ having oscillator strengths $\hsigma_j$. We introduce an effective resonance wave vector 
$\hat{q}_j$ as $\hat{q}^2_j=\Omega_j^2/c^2-p^2$ and re-write \Eq{eqn:eps22} as,
\bea \Delta\heps_\omega(\brd)&=&\Delta\heps(\brd)+\sum_j\dfrac{\Delta\hsigma_j(\brd)}{k^2-\hat{q}^2_j} \label{eqn:eps2} 
\\ &=&\Delta\heps(\brd)+\sum_j\left(\dfrac{\Delta\hsigma_j(\brd)}{2k(k-\hat{q}_j)}+\dfrac{\Delta\hsigma_j(\brd)}{2k(k+\hat{q}_j)}\right),  \nonumber \eea

In the Appendix A I use my method of \cite{DoostARX15}, which was first adapted for full dispersion of 3D nano-particles by E. A. Muljarov in Ref.\cite{Egor}, to derive for non-vanishing $\hsigma_j$ the sum rule
\be
\sum_n\frac{{\bf E_n}(\brd)\otimes {\bf E_n}(\brd ')}{2k_{n}(k_{n}\pm \hat{q}_j)}=0\,.
\label{sum_rule2}
\ee
which allows us to write the Green's function as
\be
{\hat{\mathbf{G}}^{(\pm j)}_k}(\brd,\brd ')=\sum_n\frac{{\bf E_n}(\brd)\otimes {\bf E_n}(\brd ')(k\pm \hat{q}_j)}{2k_{n}(k-k_n)(k_n\pm\hat{q}_j)}\,,
\label{GFj}
\ee
which has the useful $(k\pm \hat{q}_j)$ in the numerator for reducing the order of the perturbation matrix problem through cancellation with perturbation denominators which would otherwise have
lead to high order polynomial eigen problems.

We now consider a perturbation $\Delta\heps_{\omega}(\brd)$ of the RP inside the layer $|\brd |<a$. Similar to the previous section we use the \Eq{GFequ} to solve the perturbed problem,
\be \Ep(\brd)=-\frac{\omega^2}{c^2} \int \GFk(\brd,\brd ')\Delta\heps_{\omega}(\brd ')\Ep(\brd ') d{\bf \brd}'\,,
\label{GFsol2a} \ee
where we take $\Delta\heps_{\omega}(\brd)$ of the same form as \Eq{eqn:eps2}. Using both forms of the Green's function \cite{Egor}, \Eq{GF2} and \Eq{GFj} in \Eq{GFsol2a}, yields
\bea
\Ep(\brd)&=&-\frac{\omega^2}{c^2}\int\GFTk(\brd,\brd ')\Delta\heps(\brd ')\Ep(\brd ') d{\bf \brd}'\label{Dyson2}\\
&&-\frac{\omega^2}{c^2} \sum_{\pm, j}\int G^{(\pm j)}_k(\brd,\brd ')\dfrac{\Delta\hsigma_j(\brd')}{2k(k\pm\hat{q}_j)}\Ep(\brd ') d{\bf\brd}'\,,  \nonumber
\eea
which results in the following relationship between unperturbed and perturbed modes
\bea \Ep(\brd)&=&-\left(k^2+p^2\right)\sum_n\frac{{\bf E_n}(\brd)}{2k_n}\left[\frac{1}{k-k_n}-\dfrac{k}{k^2+p^2}\right] \nonumber\\
&&\times \int {\bf E_n}(\brd ')\Delta\heps(\brd')\Ep(\brd ')d\brd' \nonumber\\
&&-\left(k^2+p^2\right)\sum_{n,\pm , j}\frac{{\bf E_n}(\brd)(k\pm\hat{q}_j)}{2k_n(k_n\pm\hat{q}_j)(k-k_n)} \nonumber\\
&&\times\int {\bf E_n}(\brd ')\dfrac{\Delta\hsigma_j(\brd')}{2k(k\pm\hat{q}_j)}\Ep(\brd ')d\brd'\,. \label{Dyson2a} \eea
The perturbed mode $\Ep(\brd)$ satisfies \Eq{ME1D} with $\heps_{\omega}(\brd)$ replaced by $\heps_{\omega}(\brd)+\Delta\heps_{\omega}(\brd)$ and outgoing boundary conditions with the wave vector $k$.
In the interior region $\left| \brd\right|<a$ which contains the perturbation, the perturbed RSs $\Epn$ of wavenumbers $\varkappa_\nu$ can be expanded into the unperturbed ones, 
exploiting the completeness of the latter which follows from \Eq{eqn:SR5} (see also Appendix E),
\be \Epn(\brd)=\sum_{n}b_{n\nu}{\bf E_n}(\brd)\,. \label{superp} \ee
Substituting this expansion into \Eq{Dyson2a} and equating coefficients at the same basis functions ${\bf E_n}$ results in the matrix equation
\bea b_{n\nu}&=&-\frac{p^2+\varkappa_\nu k_n}{2k_n(\varkappa_\nu-k_n)}\sum_m V_{nm}b_{m\nu} \nonumber\\ 
&&-\dfrac{(\varkappa_\nu^2+p^2)}{2\varkappa_\nu(\varkappa_\nu-k_n)}\sum_{m,j} \dfrac{A^{(j)}_{nm}}{k_n^2-\hat{q}^2_j}b_{m\nu}\,,
\label{Dyson3} \eea
where
\be V_{nm}=\int{{\bf E_n}(\brd)\cdot\Delta\heps(\brd){\bf E_m}(\brd)\,d\brd}
\label{Vnm} \ee
and
\be A^{(j)}_{nm}=\int{{\bf E_n}(\brd)\cdot\Delta\hsigma_j(\brd){\bf E_m}(\brd)\,d\brd} 
\label{Ajnm} \ee
is the matrix of the perturbation in the basis of unperturbed RSs. Introducing the abbreviation
\be U_{nm}=\sum_{j}\dfrac{A^{(j)}_{nm}}{2(k^2_n-\hat{q}^2_j)} \ee
I arrive at
\begin{multline}
\sum_m b_{m\nu}\left\{ p^2U_{nm}+\varkappa_\nu\left(\dfrac{p^2V_{nm}}{2k_n}-\delta_{nm}k_m\right)\right.
\\
\left.+\varkappa_\nu^2\left(\delta_{nm}+\dfrac{V_{nm}}{2}+U_{nm}\right)\right\}=0\,,
\label{eqn:RSEquad}
\end{multline}
which is of {\it second order} in $\varkappa_\nu$. We can write this matrix problem compactly as $\mathbf{Q}(\varkappa_\nu)\mathbf{b}_\nu=0$ with $\mathbf{Q}(\varkappa_\nu)=\varkappa_\nu^2\mathbf{M}+\varkappa_\nu\mathbf{C}+\mathbf{K}$. To solve this matrix problem for a basis of size $N$, I follow Ref.\cite{TisseurBOOK2001} and use the first companion linearization, defining
\be
\mathbf{R}(\varkappa_\nu) = \varkappa_\nu\begin{bmatrix} \mathbf{M} & 0 \\ 0 & \mathbf{I} \end{bmatrix} + \begin{bmatrix} \mathbf{C} & \mathbf{K} \\ -\mathbf{I} & 0 \end{bmatrix}
\ee
where $\mathbf{I}$ is the $N$-by-$N$ identity matrix. with the corresponding vector
\be
\mathbf{z} = \begin{bmatrix} \varkappa_\nu\mathbf{b}_\nu \\ \mathbf{b}_\nu \end{bmatrix}
\ee
We solve $\mathbf{R}(\varkappa_\nu)\mathbf{z} = 0$ using the generalized eigenvalue solver from the numerical algorithms group (NAG) C++ library. We then take the first $N$ components of $\mathbf{z}$ as the eigenvector $\mathbf{b}_\nu$.

\section{Possible explanation for the difference in order between the two methods}
\label{sec:TOPOLOGY}

Here I go beyond the algebraic mathematics presented so far and discuss the fundamental topological differences between the two RSE approaches that I have developed.

The RSE perturbation theory presented here is an injective continuous mapping function except in the neighbourhood of the finite number of poles such that,
\be
\heps_{\omega}\omega^2=[\heps_{\omega}\omega^2]^{(\rm reg)}+\sum_j[\heps^j_{\omega}\omega^2]^{(\rm poles)}
\label{RESPONSE1}
\ee
is mapped to 
\bea
\heps_{\omega}\omega^2+\Delta\heps_{\omega}\omega^2&=&[\heps_{\omega}\omega^2+\Delta\heps_{\omega}\omega^2]^{(\rm reg)}\\
&&+\sum_j[\heps^j_{\omega}\omega^2+\Delta\heps^j_{\omega}\omega^2]^{(\rm poles)},\nonumber
\label{RESPONSE2}
\eea
in the following way
\be
{\bf RSE} : [\heps_{\omega}\omega^2]^{(\rm reg)}\rightarrow [\heps_{\omega}\omega^2+\Delta\heps_{\omega}\omega^2]^{(\rm reg)}
\label{RESPONSE3}
\ee
\be
{\bf RSE} : [\heps^j_{\omega}\omega^2]^{(\rm poles)}\rightarrow [\heps^j_{\omega}\omega^2+\Delta\heps^j_{\omega}\omega^2]^{(\rm poles)}
\label{RESPONSE4}
\ee
\be
{\bf RSE} : \{k_n\}\rightarrow \{\varkappa_\nu\}
\label{RESPONSE5}
\ee
\be
{\bf RSE} : \{\En\}\rightarrow\{\Epn\}
\label{RESPONSE6}
\ee
where $[\heps^j_{\omega}\omega^2]^{(\rm poles)}$ and $[\Delta\heps^j_{\omega}\omega^2]^{(\rm poles)}$ have poles in the same position in frequency space while $[\heps_{\omega}\omega^2]^{(\rm reg)}$
and $[\heps_{\omega}\omega^2+\Delta\heps_{\omega}\omega^2]^{(\rm reg)}$ are regular functions.

The mapping in \Eq{RESPONSE3} is between two spaces which are topologically equivalent to non-dispersive spaces. Therefore, the mapping problem is mathematically equivalent 
to non-dispersive RSE perturbation theory and so the order of the eigenvalue problem is not increased upon that of the non-dispersive problem.

However, in the case of mapping by \Eq{RESPONSE4} $[\heps^j_{\omega}\omega^2]^{(\rm poles)}$ contains poles in frequency space and so cannot be continuously deformed 
into $[\heps_{\omega}\omega^2]^{(\rm reg)}$ hence the map given by \Eq{RESPONSE4} is not equivalent to a non-dispersive mapping and so the order of the RSE eigenvalue problem 
is increased upon that of the non-dispersive problem.

Formally ${\bf A}$ and ${\bf B}$ are topologically equivalent if there is a homeomorphism mapping orbits of ${\bf A}$ to orbits of ${\bf B}$ homeomorphically, and preserving orientation of the orbits.

\section{Application to a planar waveguide}
\label{sec:Unperturbed}

In this section I discuss the application of the Sellmeier waveguide RSE to an effectively 1D planar waveguide system, translationally invariant in the Cartesian $z$ and $y$ directions, described by a scalar RP, i.e., $\heps_{\omega}(x)=\hat{\mathbf{1}}\varepsilon_{\omega}(x)$, $\Delta\heps_{\omega}(x)=\hat{\mathbf{1}}\Delta\varepsilon_{\omega}(x)$. As unperturbed system I use a homogeneous planar wave guide of half width $a$, so that
\be
\varepsilon_{\omega}(x)=\left\{
\begin{array}{lc}  \epsilon_{\omega}  & \mbox{for}\ \  |x|<a\,,\\
1 & \mbox{elsewhere}\,.\end{array}\right.
\ee

\subsection{Unperturbed resonant states}
The solutions of \Eq{ME1D}, which satisfy the outgoing-wave boundary conditions in TE polarization take the form \cite{ArmitagePRA14}
\be E_n(x)=\left\{
\begin{array}{lll}
(-1)^nA_ne^{-ik_nx}\,, & & x\,\leqslant-a\,,\\
B_n[e^{iq_nx}+(-1)^ne^{-iq_nx}]\,, &  &\!\!|x|\leqslant a\,,\\
A_ne^{ik_nx}\,, && x\,\geqslant a\,,
\end{array} \right.
\label{En} \ee
where the eigenvalues $k_n$ satisfy the secular equation
\be \left(k_n-q_n\right)e^{iq_na}+\left(-1\right)^n\left(k_n+q_n\right)e^{-iq_na}=0\,,
\label{secular} \ee
with 
\be q_n=\sqrt{\epsilon_{\omega} k_n^2+(\epsilon_{\omega}-1)p^2}\,.\label{eqn:secularq}\ee
I use here an integer index $n$ which takes even (odd) values for symmetric (anti-symmetric) RSs, respectively.
The normalization constants $A_n$ and $B_n$ are found from the continuity of ${\bf E_n}$ across the boundaries and the normalization condition found in Appendix F. 

In order to arrive at the normalization condition I consider an effectively 2D system and I take ${\bf\bar{E}}(\brd,k)$ as being the analytic continuation into $k$ space of ${\bf E_n(\brd)}$ and 
$A$ being an arbitrary cross-section of the translational invariant waveguide to arrive at (see Appendix F)
\bea
\label{normaliz}
1+\delta_{k_n,0}&=&\int_A{\bf E_n}(\brd)\cdot\dfrac{\partial(\omega^2\heps_\omega(\brd))}{\partial(\omega^2)}\biggr\rvert_{\omega=\omega_n}\!\!{\bf E_n}(\brd)d{\brd}\\
&&+\lim_{k\rightarrow k_n}\int_A\dfrac{{\bf E_n(\brd)}L{\bf \bar{E}(\brd,k)}-{\bf \bar{E}(\brd,k)}L{\bf E_n(\brd)}}{k^2-k^2_n}d{\brd}\nonumber
\eea
outside the system in free space Maxwell's equation simplifies (see Appendixes F and G), therefore by extending $A$ outside the system and denoting its circumference $L_A$ I can write,
\bea
\label{normaliz9}
1+\delta_{k_n,0}&=&\int_A{\bf E_n}(\brd)\cdot\dfrac{\partial(\omega^2\heps_\omega(\brd))}{\partial(\omega^2)}\biggr\rvert_{\omega=\omega_n}\!\!{\bf E_n}(\brd)d{\brd}\\
&&+\lim_{k\rightarrow k_n}\oint_{L_A} \dfrac{{\bf E_n}\cdot\nabla{\bf \bar{E}}-{\bf \bar{E}}\cdot\nabla{\bf E_n}}{k^2-k^2_n} d{\bf L}\,.\nonumber
\eea
I made the necessary assumption that $\heps_\omega$ is a (real) symmetric matrix or a scalar so that (defined to be in this case) $\bE\cdot\heps_{\omega_n}\En=\En\cdot\heps_{\omega_n}\bE$ and non dispersive at high frequencies
(see Appendix E and F).

Various schemes exist to evaluate the line integral limit in \Eq{normaliz9} such as analytic methods in Ref.\cite{MuljarovEPL10,DoostPRA14} or numerically extending the surface into a non-reflecting, absorbing, perfectly matched layer where it vanishes. 
Hence, I derive from \Eq{normaliz9} the relevant normalization condition for the planar waveguide systems, which I will use for the numerical demonstration,
\bea
\label{normaliz2}
&&\int_{-a}^{a} \dfrac{\partial(\omega^2\varepsilon_{\omega}(x))}{\partial(\omega^2)}\biggr\rvert_{\omega=\omega_n}{E}_{n} (x){E}_{n} (x)\,dx \nonumber\\
&&- \frac{{E}_{n} (-a){E}_{n} (-a) +{E}_{n} (a){E}_{n} (a)}{i2{k}_{n}}=1+\delta_{k_n,0}.
\eea
Here in \Eq{normaliz2} the first integral is taken over an arbitrary simply connected line enclosing the inhomogeneity of the system and the center of the coordinates used, and the second term is evaluated at its end points. 

The coefficients in \Eq{En}  take the form 
\be B_n=\frac{(-i)^{n}}{\sqrt{4(a\epw+ip^2/(k_n\omega^2_n/c^2))+M_n\eta}} \,, \label{ATE} \ee
\be A_n=B_n\frac{e^{iq_na}+(-1)^{n}e^{-iq_na}}{e^{ik_na}} \,, \label{BTE} \ee
with 
\be\eta=\frac{1}{\epsilon_{\omega}}\frac{\partial(\omega^2 \epsilon_{\omega})}{\partial({\omega}^2)}-1\,, \ee
and
\be M_n=\epw\left[\dfrac{(-1)^n2\sin(2q_na)}{q_n}+4a\right]\,, \ee
where $\omega_n^2/c^2=k_n^2+p^2$.

Around each pole of $\varepsilon_{\omega}$ the secular equation \Eq{secular} has a countable infinite number of solutions, each one creating a RS.  An analytic approximation to these solutions is given in Appendix B, which are used as starting values for the numerical solution of \Eq{secular} to determine these RSs closely spaced in the complex frequency plane.

\subsection{BK7 Sellmeier dispersion}

\begin{figure}
\includegraphics*[width=\columnwidth]{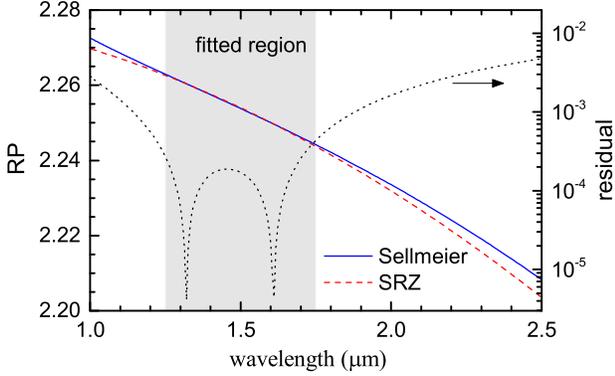}
\caption{RP $\varepsilon_{\omega}$ of SCHOTT BK7 glass given by the Sellmeier expression \Eq{eqn:BK7} (solid line), and fitted using a single resonance at zero frequency (dashed line). The absolute residual of the fit is given as dotted line.}\label{fig:BK7FitQ}
\end{figure}

In our numerical examples I use dispersion parameters describing the common borosilicate glass SCHOTT BK7. Its refractive index $n_r$ in the optical frequency range is well described by the Sellmeier expression
\bea\label{eqn:BK7}
\epsilon_{\omega}&=&1+\frac{1.03961212\lambda^2}{\lambda^2-6000.69867\,\mbox{nm}^2}\\
&&+\frac{0.231792344\lambda^2}{\lambda^2-20017.9144\,\mbox{nm}^2}+
\frac{1.01046945\lambda^2}{\lambda^2-103.560653\,\mu\mbox{m}^2}\,.
\nonumber\eea
\bea\label{eqn:BK7a}
\epsilon_{\omega}&=&1-\frac{6839.577136}{(a\omega/c)^2-6578.970180}\\
&&-\frac{457.1302872}{(a\omega/c)^2-1972.154382}-
\frac{0.3852016551}{(a\omega/c)^2-0.3812105899}
\nonumber\eea
The resulting RP is real and is shown in \Fig{fig:BK7FitQ}. This dispersion is of the form \Eq{eqn:eps} and can be treated using the quadratic matrix equation \Eq{eqn:RSEquad}. In order to compare this result with the one of the linear matrix equation \Eq{Dyson4}, I have fitted by $\epsilon_{\omega}$ over the wavelength range from $1.25\,\mu$m to $1.75\,\mu$m with the form of \Eq{del-eps}, yielding 
\be\label{eqn:BK7fit}
\epsilon_{\omega}=2.28239-0.01262\,\lambda^2/\mu\mbox{m}^2\ee
The deviations of this fit over the fitted range are around $10^{-4}$, as shown in \Fig{fig:BK7FitQ}.

\begin{figure}
	\includegraphics*[width=\columnwidth]{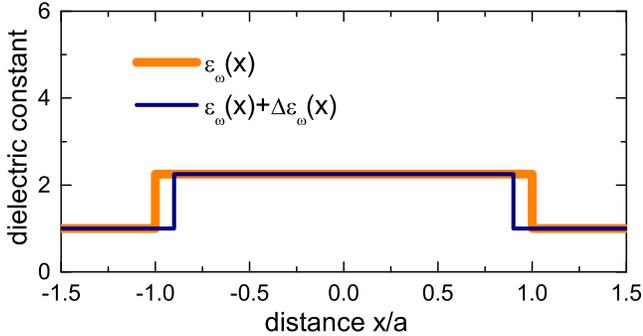}
	\caption{Cross-section of the perturbed and unperturbed wave guide used in \Sec{sec:SRZ} and \Sec{subsec:SellDis}, shown at a particular frequency.}\label{fig:WGSketch}
\end{figure}

\subsection{Unperturbed and perturbed system}
\label{subsec:system}
The unperturbed system I consider in this work as example is a planar waveguide of width $2a=2\,\mu$m. I consider a size perturbation, narrowing the waveguide by $10$ percent, as shown in \Fig{fig:WGSketch}. 
For the in-plane wave vector component I have chosen $pa=5$. I choose as basis of size $N$  for the RSE all RSs with 
\be |k_n \sqrt{\epsilon_{\omega_n}}|<\kmax(N)\,,\label{eqn:basissel}\ee
i.e., with a wave vector in the medium below a suitably chosen maximum $\kmax(N)$, this follows the approach of Ref.\cite{Egor}.

The motivation for the choice of basis selection criteria is the analogy between the RSE and a Fourier expansion. As I increase $k_n \sqrt{\epsilon_{\omega_n}}$, I increase the number of oscillations in the field as can be seen from \Eq{En}. These increasingly oscillating fields go into the basis and similar to Fourier series, increase the resolution of the composite field generated by the expansion.

\subsection{Single Sellmeier resonance at zero frequency} \label{sec:SRZ}
\begin{figure}
	\includegraphics*[width=\columnwidth]{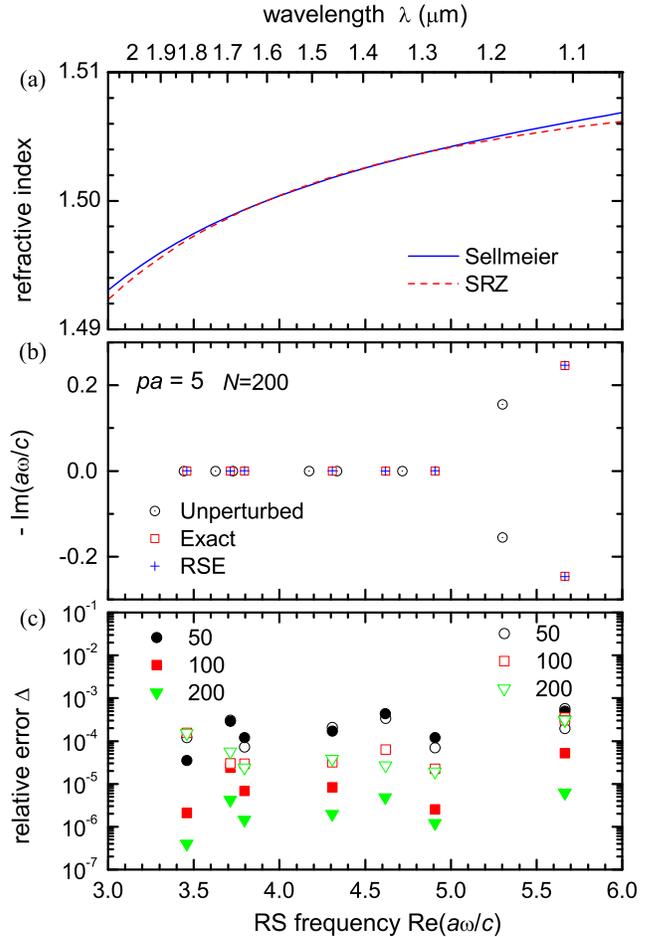}
	\caption{SRZ RSE results for a thickness perturbation of a planar waveguide as function of the real part of $\omega$. (a) Refractive index of the unperturbed and perturbed  medium (dashed line) described by the SRZ \Eq{eqn:BK7fit} and of the BK7 Sellmeier dispersion (solid line) \Eq{eqn:BK7}. (b) RS frequencies for $N=200$. Shown are exact unperturbed (open circles), exact perturbed (open squares), and RSE perturbed (crosses) data. (c) Relative error $\Delta$ of the RSE perturbed RS frequencies for $N=50,100,200$. Closed symbols give relative error to the  SRZ \Eq{eqn:BK7fit} RS eigen-frequencies as shown. Open symbols give relative error to the Sellmeier dispersion RS eigen-frequencies as shown. I show in (b) states with both incoming and outgoing boundary conditions, i.e. having positive or negative imaginary parts of complex resonant frequency, only states with outgoing boundary conditions go into the basis.}\label{fig:ToDispSRZ}
\end{figure}
Here I show results of the SRZ yielding the linear eigenvalue problem \Eq{Dyson4}. For the unperturbed and perturbed system I use a RP given by the SRZ \Eq{eqn:BK7fit}. The refractive index of the unperturbed and perturbed system is compared in \Fig{fig:ToDispSRZ} with the Sellmeier expression \Eq{eqn:BK7}. I can see that for the chosen waveguide width, the fitted wavelength range corresponds to $3.6<a\omega/c<5$. The RS frequencies are given in \Fig{fig:ToDispSRZ}(b) for the unperturbed system $\omega_n=c\sqrt{k_n^2+p^2}$ and for the perturbed system $\varpi_\nu=c\sqrt{\varkappa_\nu^2+p^2}$.
To compare $\varpi_\nu$ calculated using the RSE with the exact result $\varpi_\nu^{\rm (exact)}$ obtained by the secular \Eq{secular}, I define the relative error $\Delta=\bigl|\varpi_\nu/\varpi_\nu^{\rm (exact)}-1\bigr|$, and give the resulting values in \Fig{fig:ToDispSRZ}(c). I find that as I increase $N$, $\Delta$ decreases proportional to $N^{-3}$, similar to the findings for the non-dispersive RSE \cite{DoostPRA14, DoostPRA13, DoostPRA12}, and values in the $10^{-6}$ range are reached for $N=200$. This is actually smaller than the relative error due to the SRZ approximation of the Sellmeier dispersion (see \Fig{fig:BK7FitQ}).

\begin{figure}
	\includegraphics*[width=\columnwidth]{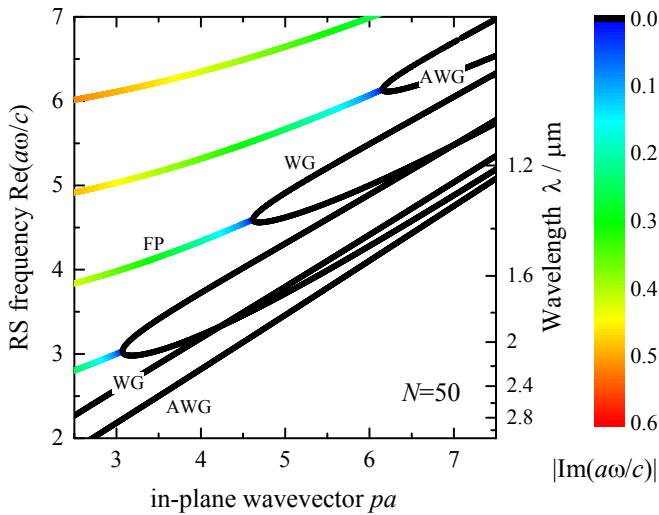}
	\caption{SRZ RSE results for a thickness perturbation of a planar waveguide and $N=50$. Shown are $\Re(\omega_\nu)$ as function of the in-plane wave vector $p$ with $\Im(\omega_\nu)$ given by the color according to the scale shown.}\label{fig:ToDispSRZ_Imag}
\end{figure}

The evolution of the perturbed RSs wavenumbers $\varpi_\nu$ with the in-plane wave vector $p$ is shown in \Fig{fig:ToDispSRZ_Imag}. I can distinguish \cite{ArmitagePRA14} the waveguide (WG) and anti-waveguide (AWG) modes, which have real $\varpi_\nu$, and the Fabry-P\'erot (FP) modes which have a finite imaginary part representing their losses. Increasing $p$, the FP modes split into WG and AWG mode at the bifurcation point at which $pc=\varpi_\nu$, at which $k=0$ i.e. at grazing incidence of the external field.

\begin{figure}
\includegraphics*[width=\columnwidth]{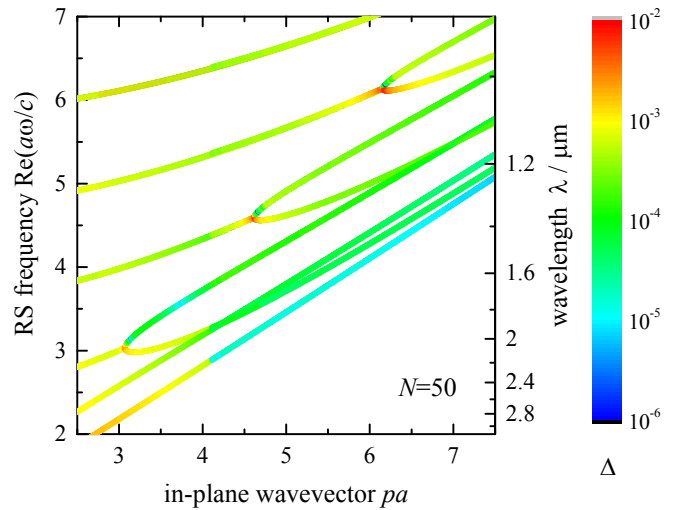}
\caption{SRZ RSE results for a thickness perturbation of a planar waveguide and $N=50$. Shown are $\Re(\omega_\nu)$ as function of the in-plane wave vector $p$ with the relative error $\Delta$ given by the color according to the scale shown.}\label{fig:ToDispSRZ_Error}
\end{figure}

The relative error $\Delta$ of $\varpi_\nu$ is given in \Fig{fig:ToDispSRZ_Error} as function of $p$. I can see that $\Delta$ has generally a weak dependence on $p$, except close to the bifurcation points, where the error of the FP and AWG mode is significantly increased. Under closer examination I see further smaller discontinuities as function of $p$, which correspond to a change of the basis states included according to \Eq{eqn:basissel} for fixed $N=50$. These results indicates that the RSE is able to reproduce all relevant RSs with a good accuracy.

\subsection{Sellmeier Dispersion}\label{subsec:SellDis}
\begin{figure}
	\includegraphics*[width=\columnwidth]{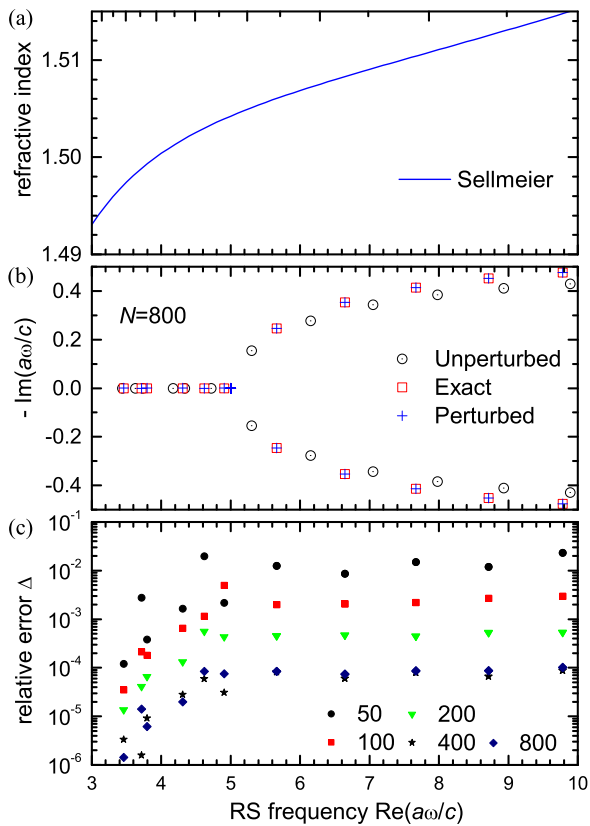}
	\caption{Sellmeier RSE results for a thickness perturbation of a planar waveguide. (a) refractive index of the Sellmeier dispersion given by \Eq{eqn:BK7}. (b) RS frequencies  for $N=800$ basis states. Shown are exact unperturbed (open circles), exact perturbed (open squares), and RSE perturbed (crosses) data. (c) relative error $\Delta$ of the RSE perturbed RS frequencies $\varpi_\nu$ for $N=50, 100, 200, 400, 800$ basis states, as labeled. I show in (b) states with both incoming and outgoing boundary conditions, i.e. having positive or negative imaginary parts of complex resonant frequency, only states with out-going boundary conditions go into the basis.}\label{fig:ToDisp}
\end{figure} 

Here I show results of the Sellmeier RSE yielding the quadratic eigenvalue problem \Eq{eqn:RSEquad}. 

The RS frequencies are given in \Fig{fig:ToDisp}(b) for the unperturbed system $\omega_n=c\sqrt{k_n^2+p^2}$ and for the perturbed system $\varpi_\nu=c\sqrt{\varkappa_\nu^2+p^2}$.
The relative error $\Delta$ is given in \Fig{fig:ToDisp}(c). Also here I find that as we increase $N$, $\Delta$ decreases proportional to $N^{-3}$, and values in the $10^{-5}$ range are reached for $N=800$.

\begin{figure}
	\includegraphics*[width=\columnwidth]{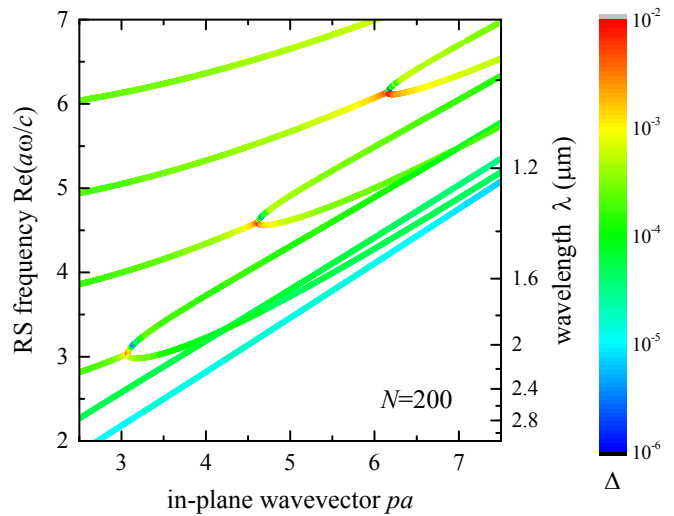}
	\caption{Sellmeier RSE results for a thickness perturbation of a planar waveguide and $N=200$. Shown are $\Re(\varpi_\nu$) as function of the in-plane wave vector $p$ with the relative error $\Delta$ given by the colour according to the scale shown.}\label{fig:ToDisp_Error}
\end{figure}

The evolution of the perturbed RSs wavenumbers $\varpi_\nu$ with the in-plane wave vector $p$ is shown in \Fig{fig:ToDisp_Error} including the relative error $\Delta$, and in \Fig{fig:ToDispSRZ_Imag} including the imaginary part $\Im( \varpi_\nu)$. 

I can see that $\Delta$ has generally a weak dependence on $p$, except for the bifurcation points, at which a Fabry-Perot mode splits into a waveguide and anti-waveguide mode.

These results indicate that with full Sellmeier dispersion the RSE is able to reproduce all relevant RSs with a good accuracy.

\subsection{Performance Comparison}\label{sec:Performance}
\begin{figure}
\includegraphics*[width=\columnwidth]{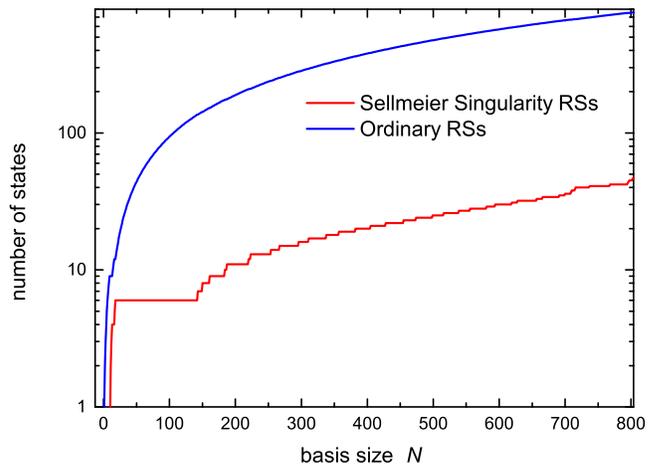}
\caption{Graph showing quantitatively how the number of poles in the basis associated closely with the poles in the Sellmeier equation for $pa=5$, those eigenfrequencies 
found using the solutions to \Eq{QUAD1} as a starting point for the Newton-Raphson search, increase with basis size $N$ when the basis is selected according to \Eq{eqn:basissel}. 
For reference I also show the number of ordinary poles, those found with the Newton-Raphson method without requiring the solutions of \Eq{QUAD1} as a starting point for their search and discovery, 
in the basis versus basis size $N$.}\label{fig:WL}
\end{figure}

Here I compare the computational complexity of the SRZ RSE and the Sellmeier RSE. The corresponding eigenvalue problems \Eq{Dyson4} and \Eq{eqn:RSEquad} show that for the same number of basis states $N$, the 
SRZ RSE is a generalized eigenvalue problem of size $N \times N$, while the Sellmeier RSE has a size $2N \times 2N$ due to the quadratic nature of \Eq{eqn:RSEquad}. Also I see from \Fig{fig:WL} that for $pa=5$ the 
resonances closely associated to the Sellmeier $\hat{q}_j$ poles, those eigen-frequencies found using the solutions to \Eq{QUAD1} as a starting point for the Newton-Raphson search, contribute approximately $10\%$ to the 
basis size thereby increasing the numerical complexity and reducing the efficiency.

\begin{figure}
\includegraphics*[width=\columnwidth]{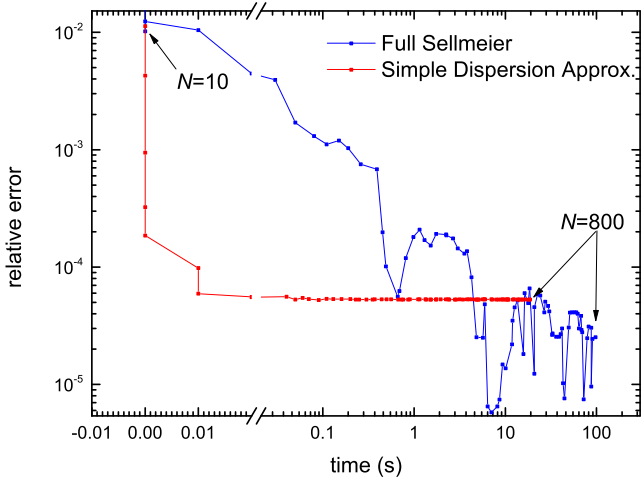}
\caption{Comparison of the average relative error of the wave guide modes occurring in the range $3<a\omega<5$ calculated using the two different perturbation schemes in this paper versus time required to diagonalise the perturbation matrix using an Intel Core 2 Duo, 6M Cache, 3.16GHz, 1333MHz FSB processor connected to 4GB of RAM and using the NAG generalised eigenvalue problem solver software}\label{fig:Performance}
\end{figure}

In \Fig{fig:Performance} I show a comparison between the average relative error in the perturbed resonances for $pa=5$ found in the range $3<a\omega<5$ versus  the number of seconds required to diagonalise the 
perturbation eigenvalue problems using an Intel Core 2 Duo, 6M Cache, 3.16GHz, 1333MHz FSB processor connected to 4GB of RAM and using the NAG generalized eigenvalue problem solver software.

For the case of SRZ dispersion the relative error reaches the limit of about $5\times 10^{-5}$ due to the RP approximation  at $N\approx 70$. Once the perturbation method using the full Sellmeier dispersion reaches this 
relative error I see that upon adding further basis states more noise in the relative error plot is generated and that the average relative error changes little. \Fig{fig:Performance} suggests that the SRZ RSE is 
several orders of magnitude more efficient than the full Sellmeier RSE when considering the resonances over a small range of frequencies such as the small range used for optical communications.     

\section{RSE Born approximation for planar waveguides}
\label{RSEBORN22}

In this section I derive the RSE Born approximation for planar waveguide, the equivalent theory for effectively 2D systems is given in Appendix G.

In effectively one dimension \Eq{ME} becomes in free space (taking $c=1$),
\be
\left[\nabla^2-\frac{1}{c^2}\frac{\partial^2}{\partial t^2}\right]\theta_p{E}({x})=0\,,
\label{ME6}
\ee
However, $\nabla^2\theta_p=-p^2\theta_p$, therefore
\be
\left[\frac{d^2}{dx^2}+k^2\right]{E}({x})=0\,,
\label{ME7}
\ee
Hence the free space GF equation is,
\be
\left[\frac{d^2}{dx^2}+k^2\right]\GFkfsOD(x,x')=\delta(x-x')\,,
 \label{GFequ0fs}
\ee
which has the solution,
\be
\GFkfsOD(x,x')=-\dfrac{e^{ik|x-x'|}}{2ik}
 \label{GFequ0fsSol1}
\ee
The systems associated with $\GFkfsOD$ and $\GFkOD$ of \Eq{GFequ} are related by the Dyson Equations perturbing back and forth with $\Delta\eps_{\omega}(x)=\eps_{\omega}(x)-1$ similar to Ref.\cite{Ge14},

\bea \GFkOD(x,x'')&=&\GFkfsOD(x,x'') \label{Dyson1}\\
&&-\omega^2\int \GFkfsOD(x,x''')\Delta\eps_{\omega}(x''')\GFkOD(x''',x'') dx'''\,, \nonumber \eea

\bea \GFkOD(x''',x'')&=&\GFkfsOD(x''',x'') \label{Dyson2}\\
&&-\omega^2\int \GFkOD(x''',x')\Delta\eps_{\omega}(x')\GFkfsOD(x',x'') dx'\,, \nonumber \eea

Combining \Eq{Dyson1} and \Eq{Dyson2} it is obtained similar to Ref.\cite{Ge14}
\begin{multline}\label{eq:Mark01}
\GFkOD(x,x'')=\GFkfsOD(x,x'')
\\
-\omega^2\int \GFkfsOD(x,x')\Delta\eps_{\omega}(x')\GFkfsOD(x',x'') dx'
\\
+\omega^4\int\int \GFkfsOD(x,x')\Delta\eps_{\omega}(x')\GFkOD(x',x''')
\\
\times\Delta\eps_{\omega}(x''')\GFkfsOD(x''',x'')dx'''dx'\,.
\end{multline}

Hence, in one dimension the RSE Born approximation can be greatly simplified by using \Eq{GFequ0fsSol1} and the spectral GF \Eq{ML2C2} in \Eq{eq:Mark01} to arrive at,
\begin{multline}\label{eq:Mark3}
\GFkOD(x,x'')=-\dfrac{e^{ik(x''-x)}}{2ik}+\dfrac{\omega^2e^{ik(x''-x)}}{4k^2}\int_{-a}^{a} \Delta\eps_{\omega}dx'
\\
-\dfrac{\omega^4 e^{ik(x''-x)}}{4k^2}\sum_n \frac{\AnOD(x)\AnOD(x'')}{2k(k-k_n)}\,.
\end{multline}
where $\AnOD$ is defined as the Fourier transform,
\be
\AnOD(x)=\int_{-a}^a e^{ik\hat{x} x'}\Delta\eps_{\omega}(x')\EnOD(x')dx'
 \label{Anic}
\ee

Interestingly in one dimension I do not require the far field approximation to make the simplification of the Green's function required to bring the RSE Born approximation to the form of \Eq{eq:Mark1}. Hence, in 
one dimension the RSE Born Approximation is valid everywhere outside of the slab and not just in the far field. I note that fast Fourier transforms are available for use upon \Eq{Anic}.

\begin{figure}
\includegraphics*[width=\columnwidth]{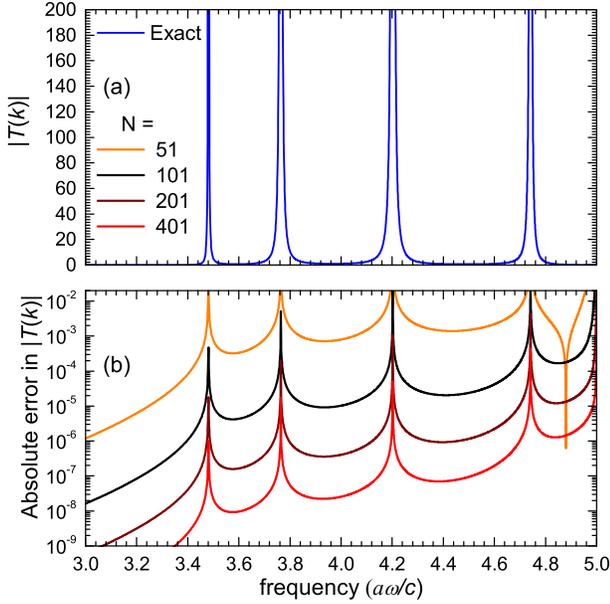}
\caption{I give transmission results for the unperturbed system in \Sec{sec:SRZ} with $pa=5$. I have used analytic modes as RSs for the RSE Born approximation. 
(a) Exact transmission as a function of frequency as defined by \Eq{Trans1} (b) Absolute error in transmission calculated using the analytic form of $T(k,x')$
between $x=-a$ and $x=a$ versus frequency $\omega$ as comparison for the RSE 
Born approximation. Here $N=51, 101, 201, 401$ as labeled.} \label{fig:RSEBORN}
\end{figure}

I demonstrate the computational accuracy of the RSE Born approximation in \Fig{fig:RSEBORN} where I calculate the transmission defined as
\begin{equation}
T(k,x')=\left|2kG(x',-a;k)\right|^{2} \label{Trans1}
\end{equation}
For comparison the analytic GF is found by solving Maxwell's wave equation in one dimension with a source of plane waves while making use of Maxwell's boundary conditions. 
The system treated is the unperturbed system in  \Sec{sec:SRZ} with $pa=5$.
From \Fig{fig:RSEBORN} we can see that unlike the standard Born approximation the RSE Born approximation is valid over an arbitrarily wide range of $k$ depending only on the basis size $N$ used. 
Furthermore we see that as the basis size increases the RSE Born approximation converges to the exact solution. The absolute error in the RSE Born approximation is approximately reduced by an order 
of magnitude each time the basis size is doubled. Absolute errors of $10^{-8}-10^{-5}$ are seen in the $k$ range shown for basis size $N=401$.
 
\section{Summary}\label{sec:summary}
In this work I have extended the RSE to media having a simple dispersion linear in wavelength squared. This dispersion has a single pole at zero frequency and does not introduce an additional dynamic degree of freedom 
as it would be the case for a more general Sellmeier model of the material response. This property allows to keep the simplicity of the RSE formulation, therefore retaining the advantage of the RSE in computational 
efficiency discussed in \Onlinecite{DoostPRA14}.

Furthermore, in this work I have also extended the RSE to waveguides obeying the Sellmeier equation for glasses. In order to do this I have reduced the order of the eigenvalue problem to second order by using sum rules
as was done in \cite{Egor}.

To make use of the RSs generated by the RSE I derived the RSE Born approximation for effectively 1D and 2D waveguides.

My conclusion is that the simple dispersion treatment is more efficient than using the full Sellmeier dispersion. This is because the dispersion only has to be correctly reproduced over a narrow part of the optical 
frequency range and therefore it is inefficient to include the full and at some frequencies unphysical Sellmeier dispersion.
 
\acknowledgments 
I acknowledge support by the Cardiff University EPSRC Doctoral Prize Fellowship EP/M50631X/1.
I thank E. A. Muljarov for his positive and highly valuable contribution to this paper, which was to present me with his copy of his quite complete
draft of Ref.\cite{Egor}. I thank W. Langbein and T. Wood for their positive contributions to this paper during the early stages.

\appendix
\section{Sum rule}
\label{App:SR}
Here I use my approach of Ref.\cite{DoostARX15}, which was first adapted for full dispersion of 3D nano-particles by E. A. Muljarov in Ref.\cite{Egor}, to derive sum rules and completeness of modes for the waveguide GF.

By substituting \Eq{GF1} into \Eq{GFequ} and using \Eq{ME1D} I have,
\be -\hat{\mathbf{1}}\delta(\brd - \brd ') = \sum_n{\dfrac{(\omega^2_n\heps_{\omega_n}-\omega^2\heps_{\omega})}{c^2 2k_n(k-k_n)}{\bf E_{n}}(\brd)\otimes {\bf E_{n}}(\brd ')}\,.\label{eqn:SR1}\ee
I now consider the pole in $\heps_{\omega}$ at $k=\pm\hat{q}_j$  
\be \heps_{\omega}=\dfrac{\hsigma_j(\brd)}{2k(k\pm\hat{q}_j)} \label{eqn:SR2}\ee
where I have been able to ignored other terms in the dispersion since they are constant in the limit $k\rightarrow\mp\hat{q}_j$.
Using $\omega^2/c^2=k^2+p^2$ and $\omega_n^2/c^2=k^2_n+p^2$ and \Eq{eqn:SR2} in \Eq{eqn:SR1}, I find
\begin{multline}
=\dfrac{(\omega^2_n\heps_{\omega_n}-\omega^2\heps_{\omega})}{c^2 2k_n(k-k_n)}
\\
={2\hsigma_j(\brd)}\dfrac{kp^2}{8k^2_nk(k-k_n)(k_n\pm\hat{q}_j)}
\\
-\dfrac{2\hsigma_j(\brd)}{(k\pm\hat{q}_j)}\left(\dfrac{p^2k_n}{8k^2_nk(k-k_n)}\pm\dfrac{k_n\hat{q}_jk}{8k^2_nk(k_n\pm\hat{q}_j)}\right)
\end{multline}
Then convoluting \Eq{eqn:SR1} with arbitrary finite function and taking the limit $k\rightarrow\mp\hat{q}_j$, I see that the second term is diverging unless
\be 
\sum_n\frac{{\bf E_{n}}(\brd)\otimes {\bf E_{n}}(\brd ')}{k_{n}(k_{n}\pm \hat{q}_j)}=0\,. 
\label{eqn:SR4}
\ee
since when $k\rightarrow\mp\hat{q}_j$
\begin{multline}
\left(\dfrac{p^2k_n}{8k^2_nk(k-k_n)}\pm\dfrac{k_n\hat{q}_jk}{8k^2_nk(k_n\pm\hat{q}_j)}\right)
\\
=\pm\dfrac{p^2+\hat{q}^2_j}{8k_n(k_n\pm\hat{q}_j)\hat{q}_j}
%{8k_n\hat{q}^_j(k_n\pm\hat{q}_j)}
\end{multline}
The closure and over completeness follow from letting $\omega\rightarrow\infty$ to give \cite{DoostARX15} 
\be 
\dfrac{\heps(\brd)}{2}\sum_n\En(\brd)\otimes\En(\brd')=\hat{\mathbf{
1}}\delta(\brd-\brd')\,, \label{eqn:SR5}
\ee

\section{Analytic solution of secular equation close to RP poles}
\label{App:IP}

Here I calculate an analytic approximation for the solutions of the secular equation \Eq{secular} close to the poles of the RP, following the approach of Ref.\cite{Egor}.
In the limit $\epsilon_{\omega}\rightarrow\infty$ I find from \Eq{eqn:secularq} that $q_n\rightarrow \sqrt{\epsilon_{\omega}}\omega_n/c$, so that \Eq{secular} is given by $-e^{iq_na}+(-1)^ne^{-iq_na}=0$, and thus
\be 
\begin{array}{lll}
	\sin(\sqrt{\epsilon_{\omega}}\omega_n a/c)=0 & \mbox{for even}\,& n\\
	\cos(\sqrt{\epsilon_{\omega}}\omega_n a/c)=0\,& \mbox{for odd}\,& n
\end{array} 
\label{eqn:IP1} \ee
In the limit $k_n\rightarrow\pm\hat{q}_j$ to the poles of the RP given in \Eq{eqn:eps2} I find 
\be
\epsilon_\omega \frac{\omega_n^2}{c^2}=\sigma_j\dfrac{k^2_n+p^2}{2k_n(k_n\mp\hat{q}_j)}=\gamma_n
\label{eqn:IP2}\ee
and solutions of \Eq{eqn:IP1} are given by 
\be \gamma_n a^2=(\pi n/2)^2
\label{IP3} \ee
The solutions of \Eq{eqn:IP2} with $k_n$ close to $\pm\hat{q}_j$ are given by the analytic solution of the quadratic formula
\be
(\sigma_j-2\gamma_n)k^2_n\pm 2\gamma_n\hat{q}_j k_n+\sigma_j p^2=0 \label{QUAD1}
\ee
I use the solutions of \Eq{QUAD1} as starting points for the Newton-Raphson search for RSs.

\begin{figure}
\includegraphics*[width=\columnwidth]{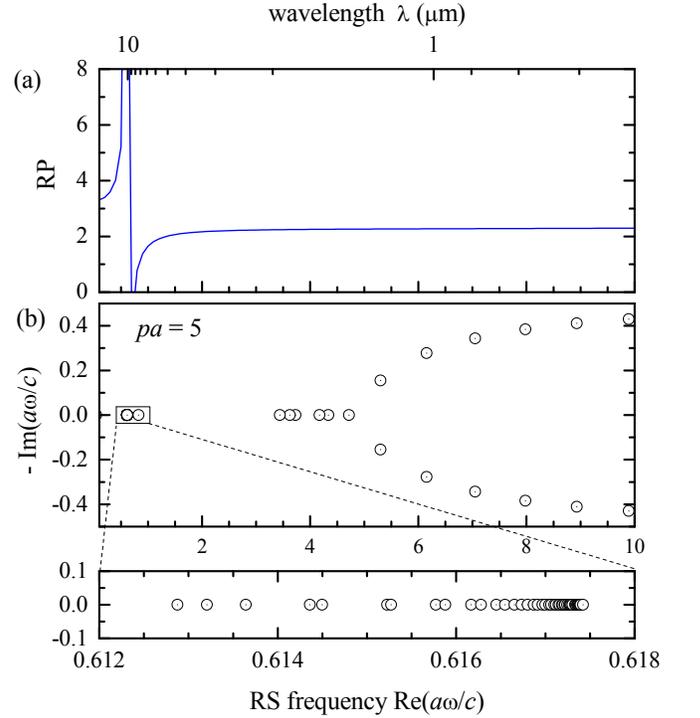}
\caption{(a) RP as a function of frequency given by \Eq{eqn:BK7}. (b) Resonant frequencies $\omega_n$ of the homogeneous dielectric waveguide with $pa=5$ and this RP, forming the basis states for the Sellmeier RSE discussed in \Sec{subsec:SellDis}. (c) Zoom of (b) showing the series of poles on the low frequency side of the singularity in RP. I show in (b) and (c) states with both incoming and outgoing boundary conditions, i.e. having positive or negative imaginary parts of complex resonant frequency, only states with outgoing boundary conditions go into the basis.}\label{fig:Infinite_poles}
\end{figure}

The RSs of the unperturbed system (see \Sec{subsec:system}) with Sellmeier dispersion \Eq{eqn:BK7} are given in \Fig{fig:Infinite_poles}. I note that close to the lowest resonance frequency of the dispersion at $\Omega_j a/c\approx 0.6174$, there are a large number of RSs approaching the pole from smaller frequencies. These RSs arise because the refractive index is diverging to positive infinity on the low frequency side, allowing for a countable infinite number of WG and AWG modes to form. $\sigma_j$ is negative for all the resonances in \Eq{eqn:BK7}.

\section{Calculation of L}
\label{App:L}

Here I derive the form of the $L$ operator occurring in the reduced Maxwell's wave equation for waveguide \Eq{ME1D}.

For planar systems $L$ is given in \Onlinecite{ArmitagePRA14} as
\be
-L=\frac{d^2}{dx^2}-p^2 \label{EF1}
\ee
For effectively two-dimensional systems with one direction of translational invariance $L$ can be calculated as follows. Due to translational invariance in the direction $z$, the direction of propagation I can write 
the full field as,
\be {\bf \hat{E}}({\bf r})=e^{ipz-i\omega t}{\bf E}(\brd)=\theta_p{\bf E}(\brd)\,,
\label{EF2} \ee
where $\brd$ give the coordinates in the plane perpendicular to $z$.

Therefore using standard vector identities I can write
\bea
\nabla\times\nabla\times{\bf \hat{E}}&=&\nabla\times\nabla\times(\theta_p{\bf E})\nonumber\\
&=&\nabla\times(\nabla\theta_p\times {\bf E}+\theta_p\nabla\times {\bf E})
\nonumber\\
&=&\nabla\theta_p(\nabla\cdot {\bf E})-{\bf E}(\nabla\cdot(\nabla\theta_p))\nonumber\\
&&+({\bf E}\cdot\nabla)\nabla\theta_p-(\nabla\theta_p\cdot\nabla){\bf E}\nonumber\\
&&+\nabla\theta_p\times\nabla\times {\bf E}\nonumber\\
&&+\theta_p\nabla\times\nabla\times {\bf E}\label{L1}
\eea
I make use of the following identities,
\bea
\nabla\theta_p&=&ip\theta_pe_z\\
\nonumber\\
\nabla\cdot\nabla\theta_p&=&-p^2\theta_p\\
\nonumber\\
\nabla\theta_p(\nabla\cdot{\bf E})&=&ip\theta_p(\nabla\cdot{\bf E})e_z\\
\nonumber\\
({\bf E}\cdot\nabla)\nabla\theta_p&=&(e_z\cdot{\bf E})p^2\theta_pe_z\\
\nonumber\\
\partial_z{\bf E}&=&0
\eea
where $e_z$ is the unit vector in the $z$ direction so as to simplify \Eq{L1} to
\bea
\nabla\times\nabla\times{\bf\hat{E}}&=&\nabla\times\nabla\times({\bf E}\theta_p)\\
&=&ip\theta_pe_z(\nabla\cdot{\bf E})+{\bf E}p^2\theta_p
\nonumber\\
&&+(e_z\cdot{\bf E})p^2\theta_p e_z-ip\theta_p\partial_z{\bf E}
\nonumber\\
&&+ip\theta_pe_z\times\nabla\times{\bf E}+\theta_p\nabla\times\nabla\times{\bf E}\nonumber\\
&=&\theta_p L({\bf E})\nonumber\\
\nonumber \eea
Hence I see that,
\bea
\nonumber\\
L({\bf E})&=&ipe_z(\nabla\cdot {\bf E})+{\bf E} p^2
\\
&&+(e_z\cdot{\bf E})p^2 e_z+ipe_z\times\nabla\times{\bf E}
\nonumber\\
&&+\nabla\times\nabla\times{\bf E}\nonumber
\eea

So $L$ is a linear operator independent of the RP and $z$ as required. Therefore the eigenvalue problems derived in \Sec{sec:SingleSell} and \Sec{sec:GenSel} are valid for planar and cylindrical wave guides. Also the derivation of $L$ demonstrates I can use separation of variables in \Eq{ME} to arrive at \Eq{EF2}.

\section{Spectral representation of the GF of an open system}
\label{SRG}

Here I almost exactly repeat my derivations which I contributed to Ref.\cite{DoostPRA13} using exactly the same method but with increased mathematical rigour, in order to prove 
in this section the spectral representation of the Green's function (GF)  of a general wave equations.

The GF of an open  waveguide equation system is a tensor function $\GFk$ which satisfies the outgoing wave BCs and the waveguide wave equation \Eq{ME1D}
with a delta function source term,
\be
-L\GFk(\brd,\brd')+\heps(\brd,k)k^2\GFk(\brd,\brd')=\hat{\mathbf{1}}\delta(\brd-\brd')\,,
 \label{GFequA}
\ee
In this Appendix the effective permittivity $\heps(\brd,k)=\heps_{\omega}(\brd)(1+{p^2}/{k^2})$ with fixed $p$. Assuming a simple-pole structure of the GF inside the scatterer with poles at $k=q_n$ and taking into account its large-$k$ vanishing asymptotics, the Mittag-Leffler  theorem allows us to 
express the GF only inside the scatterer as the convergent 
\be
\GFk(\brd,\brd')=\sum_n\frac{\Qn(\brd,\brd')}{k-q_n}\,.
 \label{ML1}
\ee
It will probably be that \Eq{ML1} will need to be experimentally verified by comparing scattering predicted by the RSE Born approximation with experimental results of scattering from well defined scatterers.
It may be that \Eq{ML1} is a fundamental law of Physics. My justification for this form of the GF is the superposition of Lorentzians which make up the scattering profile of resonators and the numerical verification 
of this form of GF made in Ref.\cite{DoostPRA12,DoostPRA13,ArmitagePRA14,DoostPRA14}.

Assuming no degeneracy with the mode $n$, the definition of the residue tensor $\Qn(\brd,\brd')$ at a simple pole of the function $\GFk(\brd,\brd')$ is,
\be
\lim_{k\to q_n}(k-q_n)\GFk(\brd,\brd')=\Qn(\brd,\brd')
\label{APRIL}
\ee
I have again assumed $\GFk(\brd,\brd')$ to be holomorphic in this neighbourhood of $q_n$ except for at the poles $q_n$ so that it has a Laurent series at $q_n$. Substituting the expression 
\Eq{ML1} into \Eq{APRIL} gives
\be
\lim_{k\to q_n}(k-q_n)\sum_m\frac{\Qm(\brd,\brd')}{k-q_m}=\Qn(\brd,\brd')
\ee
so that
\be
\lim_{k\to q_n}(k-q_n)\sum_{m\neq n}\frac{\Qm(\brd,\brd')}{k-q_m}=0
\label{AF}
\ee 
Substituting the expression \Eq{ML1} into \Eq{GFequA} and convoluting with an arbitrary finite vector function
${\bf D}(\brd)$ over a finite volume $V$ we obtain
\be
\sum_n \frac{-L\Fn({\brd})+\heps(\brd,k)k^2\Fn(\brd)}{k-q_n}={\bf D}(\brd)\,,
\label{Trainm1}
\ee
where $\Fn(\brd)=\int_V \Qn(\brd,\brd'){\bf D}(\brd')d\brd'$. Multiplying by $(k-q_n)$ and taking the limit $k\to q_n$ yields
\bea
\lim_{k\to q_n}(k-q_n)\sum_m \frac{-L\Fm({\brd})+\heps(\brd,k)k^2\Fm(\brd)}{k-q_m}
\\
=\lim_{k\to q_n}(k-q_n){\bf D}(\brd)=0\,.
\label{EW}
\eea
From \Eq{AF} we can see,
\be
\lim_{k\to q_n}(k-q_n)\sum_{m\neq n}\frac{-L\Fm({\brd})+\heps(\brd,k)k^2\Fm(\brd)}{k-q_m}=0\,,
\ee
so we can drop terms $n\neq m$ from the summation in \Eq{EW} to give
\be
 \lim_{k\to q_n}(k-q_n)\dfrac{-L\Fn(\brd)+\heps(\brd,k)k^2\Fn(\brd)}{k-q_n}=0\,.
\label{Train0}
\ee
or
\be
-L\Fn(\brd)+\heps(\brd,q_n)q^2_n\Fn(\brd)=0\,.
\label{Train0}
\ee

Due to the convolution with the GF, $\Fn(\brd)$ satisfies the same outgoing wave BCs. Then, according to \Eq{ME1D},   $\Fn(\brd)\propto \En(\brd)$ and  $q_n=k_n$, i.e.
\be
\label{me3D0}
L\En(\brd)=\heps(\brd,k_n)k^2_n\En(\brd)\,,
\ee

Note that the convolution of the kernel $\Qn(\brd,\brd')$ with different vector functions $\bf D(\brd)$ can be proportional to one and the same vector function $\En(\brd)$ only if the kernel has the form of a product:
\be \Qn(\brd,\brd')=\En(\brd)\otimes\En(\brd')/2k_n\,, \label{QnE}\ee
where $\otimes$ is the dyadic product operator.

The symmetry in \Eq{QnE} follows from the reciprocity theorem, described mathematically by the relation
 \be
{\bf s}_1\GFk(\brd_1,\brd_2) {\bf s}_2={s}_2\GFk(\brd_2,\brd_1) {s}_1\,,
 \label{reciprocity}
\ee which holds for any two point sources ${s}_{1,2}$ at points $\brd_{1,2}$ emitting at the same frequency. Hence $\GFk(\brd,\brd')$ is symmetric.

In the case of a GF made up of degenerate modes the proof of \Eq{QnE} is modified by making use of orthogonality of the degenerate modes to choose ${\bf D}({\brd})$ such that,
\be
\int_V\Em(\brd)\cdot{\bf D}({\brd})\,d\brd=0\,,
\label{sigma}
\ee
for $m\neq n$ and where state $m$ is degenerate with $n$.

Hence I obtain
\be \GFk(\brd,\brd')=\sum_n \frac{\En(\brd)\otimes\En(\brd')}{2k_n(k-k_n)}\,.
\label{TUES_NIGHT1} \ee

\section{Derivation of sum rule and completeness}
\label{App:RSEBORN101}

Here I exactly repeat my derivation of the sum rule and completeness of the GF which I made for Ref.\cite{DoostARX15} but in more detail.

In order to simplify the RSE Born approximation we require an appropriate spectral form of the GF which is different from the one already proven in Appendix D. To obtain this 
correct form I start with the GF valid inside the scatterer only,

\be \GFk(\brd,\brd')=\sum_n \frac{\En(\brd)\otimes\En(\brd')}{2k_n(k-k_n)}\,.
\label{TUES_NIGHT1} \ee
Substituting \Eq{TUES_NIGHT1} in 
\be
- L\GFk(\brd,\brd')+(k^2+p^2)\heps_{\omega}(\brd)\GFk(\brd,\brd')=\hat{\mathbf{1}}\delta(\brd-\brd')\,,
 \label{GFequ0}
\ee
gives for $k\to \infty$,
\be 
\heps(\brd)\sum_n\dfrac{(k+k_n)\En(\brd)\otimes\En(\brd')}{2k_n}=\hat{\mathbf{1}}\delta(\brd-\brd')\,. 
\label{TUES_NIGHT2A} 
\ee
since throughout the derivation in this appendix we are considering the limit where $k\to \infty$ at which $\heps_k(\brd)=\heps(\brd)$, i.e. the system is non-dispersive at high frequencies.

Convoluting \Eq{TUES_NIGHT2A} with arbitrary finite functions ${\bf D}$ which are only non-zero inside the resonator,
gives 
\be 
\int\left(\heps(\brd)\sum_n\dfrac{(k+k_n)\En(\brd)\otimes\En(\brd')}{2k_n}\right){\bf D}(\brd')d\brd'={\bf D}(\brd)\,. 
\label{TUES_NIGHT2} 
\ee
and assuming the series are convergent as $k\rightarrow\infty$ we get,
\be \int\left(\sum_n\dfrac{\En(\brd)\otimes\En(\brd')}{2k_n}\right){\bf D}(\brd')d\brd'=0\,. \label{TUES_NIGHT77} \ee
and since this is true for all ${\bf D}$ it follows that
\be \sum_n \dfrac{\En(\brd)\otimes\En(\brd')}{2k_n}=0\,. \label{Sum-rule} \ee
Combining \Eq{TUES_NIGHT1} and \Eq{Sum-rule} yields
\be \GFk(\brd,\brd')=\sum_n \frac{\En(\brd)\otimes\En(\brd')}{2k(k-k_n)}\,.
\label{ML2C2} \ee
Combining \Eq{TUES_NIGHT2A} and \Eq{Sum-rule} leads to the closure
relation
\be \dfrac{\heps(\brd)}{2}\sum_n\En(\brd)\otimes\En(\brd')=\hat{\mathbf{
1}}\delta(\brd-\brd')\,, \label{Closure} \ee
which expresses the completeness of the RSs, so that any function can be written
as a superposition of RSs. If in the perturbed system some of the series are not convergent or are instead conditionally convergent then we will not arrive at the sum rule and completeness, 
in which case I expect that the RSE Born approximation will still give 
convergence to the exact solution but only if a valid spectral Green's function is used, such as \Eq{TUES_NIGHT1}. 

\section{Derivation of the correct normalisation for effectively 2D waveguides}
\label{App:RSEBORN}

Here I almost exactly repeat my derivations which I contributed to Ref.\cite{DoostPRA14} using exactly the same method as for Maxwell's equations in order to prove in this section that the spectral 
representation, only valid inside the waveguide (similar to Ref.\cite{DoostARX15} for nano-particles), rigorously derived in Appendix C, D, and E 
\be 
\GFk(\brd,\brd')=\sum_n \frac{\En(\brd)\otimes\En(\brd')}{2k(k-k_n)}\,,
\label{doh}
\ee 
leads to the RS normalization condition \Eq{normaliz} for general waveguide equations. To do so, I consider an analytic continuation $\bE(\brd,k)$ of the vector wave function $\En(\brd)$ 
around the point $k=k_n$ in the complex $k$-plane ($k_n$ is the wavenumber of the given RS). I select the analytic continuation such that it satisfies the outgoing wave boundary condition and 
my waveguide equation (taking $c=1$)
\be
-L\bE(\brd,k)+\heps(\brd,\omega)(k^2+p^2)\bE(\brd,k)=(k^2-k_n^2)\bs(\brd)
 \label{MEcont}
\ee
with an arbitrary  source term.

The source $\bs(\brd)$ has to be zero outside the cross-section of the inhomogeneity of $\heps_{\omega}(\brd)$ for the field $\bE(\brd,k)$ to satisfy the outgoing wave boundary condition. It also has to be non-zero 
somewhere inside that cross-section, as otherwise $\bE(\brd,k)$ would be identical to $\En(\brd)$. It is further require that $\bs(\brd)$ is normalized according to
\be
\int_A\En(\brd)\cdot\bs(\brd)\,d\brd=1 + \delta_{k_n,0}\,,
\label{sigma}
\ee
The integral in \Eq{sigma} is taken over an arbitrary cross-section $A$ which includes all system inhomogeneity of $\heps_{\omega}({\brd})$. I do not derive \Eq{sigma}, it is simply a convenient condition to put on the
otherwise arbitrary spatial dependence of the source which lies only inside the waveguide scatterer. 
If I had made the condition anything else (which in my earliest proof I did) then algebra and cancellation would lead us back to the same result, however with more mathematical complexity and operations. The 
condition make the mathematics easier because it causes $\bE\rightarrow\En$ exactly and without proportionality constant appearing, equation~(\ref{sigma}) ensures that the analytic continuation reproduces 
$\En(\brd)$ in the limit $k\to k_n$. Solving \Eq{MEcont} with the help of the GF and using its spectral representation \Eq{doh}, I find:
\bea
&&\!\!\!\!\!\!\!\!\!\!\!\!\!\!\!\!\!\!\bE(\brd,k)=\int_{A} \GFk(\brd,\brd') (k^2-k_n^2)\bs(\brd') d\brd'\nonumber
\\
&&\!\!\!\!\!\!\!=\sum_{m}\Em(\brd) \frac{k^2-k_n^2}{2k(k-k_{m})} \int_{A} \Em(\brd')\cdot\bs(\brd') \,d\brd',
\label{WHY}\eea
and then, using \Eq{sigma}, obtain
$$
\lim_{k\to {k_n}} \bE(\brd,k)=\En(\brd)\,,
$$
for any $\brd$ inside the system. Outside the system, the analytic continuation $\bE(\brd,k)$ is defined as a solution of the waveguide equation wave equation in free space. This solution is connected to 
the field inside the system [given by \Eq{WHY}] through the boundary conditions. Note that in the case of degenerate modes, $k_m=k_n$ for $m\neq n$, the current $\bs(\brd)$ has to be chosen in such a 
way that it satisfies \Eq{sigma} and, additionally,
$$
\int_A{\Em}(\brd)\cdot\bs(\brd)\,d\brd=0\,,
$$
in order that the degenerate modes can be normalised separately.

I now consider the integral
\be
I_n(k) =\frac{\int_A(\bE\cdot L\En -\En\cdot L\bE)d\brd}{k^2-k_n^2}
\label{I0}
\ee
and evaluate it by using the waveguide equations Eqs.\,(\ref{ME1D}) and (\ref{MEcont}) for $\En$ and $\bE$, respectively, and the source
term normalization \Eq{sigma}:
\be
I_n(k) =\frac{\int_A (\bE\cdot \omega_n^2\heps_{\omega_n}\En -\En\cdot \omega^2\heps_{\omega}\bE)d\brd}{k^2-k_n^2} + 1 + \delta_{k_n,0}\,,
%\int_A \bE(k,\brd)\cdot\heps(\brd)\En(\brd)d\brd
\label{I1} \ee
where I assume that $\heps_\omega$ is a a (real) symmetric matrix or a scalar so that (defined to be in this case) $\bE\cdot\heps_{\omega_n}\En=\En\cdot\heps_{\omega_n}\bE$, and so I obtain also by commutation of $\bE$ and $\En$ and simple calculus in the integral shown \Eq{I1} the dispersion factor and normalization 
becomes,
\bea
\label{normaliz17}
1+\delta_{k_n,0}&=&\int_A{\bf E_n}(\brd)\cdot\dfrac{\partial(\omega^2\heps_\omega(\brd))}{\partial(\omega^2)}\biggr\rvert_{\omega=\omega_n}\!\!{\bf E_n}(\brd)d{\brd}\\
&&+\lim_{k\rightarrow k_n}\int_A\dfrac{{\bf E_n(\brd)}L{\bf \bE(\brd,k)}-{\bf \bE(\brd,k)}L{\bf E_n(\brd)}}{k^2-k^2_n}d{\brd}\nonumber
\eea
and then finally with $L_A$ the circumference of $A$,
\bea
\label{normaliz}
1+\delta_{k_n,0}&=&\int_A\En(\brd)\cdot\dfrac{\partial \omega^2\heps(\brd,\omega)}{\partial \omega^2}\bigg|_{\omega=\omega_n}\!\!\En(\brd)d{\brd}\\
&&+\lim_{k\rightarrow k_n}\oint_{L_A} \dfrac{\En\cdot\nabla\bE-\bE\cdot\nabla\En}{k^2-k^2_n} d{\bf L}\,.\nonumber
\eea
I have also assumed a simpler form for the waveguide equation in free space, see Appendix G so that I can use the vector and scalar Green's theorem to simplify \Eq{normaliz17}. Various schemes exist to evaluate the line integral limit in \Eq{normaliz} such as analytic methods in 
Ref.\cite{MuljarovEPL10} or numerically extending the closed arc into a non-reflecting, absorbing, perfectly matched layer where it vanishes.

Due to the over completeness of the basis demonstrated in Appendix E we do not have the relation we find for Hermitian systems
\be
\label{nearlydone}
\int_{\it all space}\dfrac{\omega^2_n\heps_{\omega_n}(\brd)-\omega^2_m\heps_{\omega_m}(\brd)}{\omega^2_m-\omega^2_n}{\Em}(\brd)\cdot{\En}(\brd)\,d\brd=0\,,
\ee
for all $m\neq n$. The basis states of non-Hermitian systems are not all mutually orthogonal although some are orthogonal to one another. Hence we cannot derive for non-Hermitian systems the normalization 
relation, which is well known for Hermitian systems and follows from \Eq{nearlydone}, 
\be
\label{nearlydone2}
\int_{\it all space}\dfrac{\partial(\omega^2\heps_\omega(\brd))}{\partial(\omega^2)}\biggr\rvert_{\omega=\omega_n}\!\!{\En}(\brd)\cdot{\En}(\brd)\,d\brd=1+\delta_{k_n,0}\,,
\ee
and, hence, \Eq{normaliz} and \Eq{nearlydone2} are different.

\section{RSE Born Approximation for effectively 2D waveguides}
\label{App:RSEBORN}

In effectively two dimensions \Eq{ME} becomes in free space (taking $c=1$),
\be
\left[\nabla^2-\frac{1}{c^2}\frac{\partial^2}{\partial t^2}\right]\theta_p{\bf E}(\brd)=0\,,
\label{ME6}
\ee
However $\nabla^2\theta_p=-p^2\theta_p$, therefore
\be
\left[\nabla^2+k^2\right]{\bf E}=0\,,
\label{ME7}
\ee
Hence the free space GF equation is,
\be
\left[\nabla^2+k^2\right]\GFkfs(\brd,\brd')=\hat{\mathbf{1}}\delta(\brd-\brd')\,,
 \label{GFequ0fs}
\ee
which has the solution, also assuming $\rho>>\rho'$,
\be
\GFkfs(\brd,\brd')=-\dfrac{i}{4}H_0(k|\brd-\brd'|)\hat{\mathbf{1}}\approx Q\sqrt{\dfrac{1}{\rho\pi}}e^{ik|\brd-\brd'|}\hat{\mathbf{1}} 
 \label{GFequ0fsSol}
\ee
The systems associated with $\GFkfsOD$ and $\GFkOD$ of \Eq{GFequ} are related by the Dyson Equations perturbing back and forth with $\Delta\heps_{\omega}(\brd)=\heps_{\omega}(\brd)-\hat{\mathbf{1}}$ similar to Ref.\cite{Ge14},

\bea \GFk(\brd,\brd'')&=&\GFkfs(\brd,\brd'') \label{Dyson11}\\
&&-\omega^2\int \GFkfs(\brd,\brd''')\Delta\heps_{\omega}(\brd''')\GFk(\brd''',\brd'') d\brd'''\,, \nonumber \eea

\bea \GFk(\brd''',\brd'')&=&\GFkfs(\brd''',\brd'') \label{Dyson21}\\
&&-\omega^2\int \GFk(\brd''',\brd')\Delta\heps_{\omega}(\br')\GFkfs(\brd',\brd'') d\brd'\,, \nonumber \eea

Combining \Eq{Dyson11} and \Eq{Dyson21} it is obtained similar to Ref.\cite{Ge14}
\begin{multline}\label{eq:Mark0}
\GFk(\brd,\brd'')=\GFkfs(\brd,\brd'')
\\
-\omega^2\int \GFkfs(\brd,\brd')\Delta\heps_{\omega}(\brd')\GFkfs(\brd',\brd'') d\brd'
\\
+\omega^4\int\int \GFkfs(\brd,\brd')\Delta\heps_{\omega}(\brd')\GFk(\brd',\brd''')
\\
\times\Delta\heps_{\omega}(\brd''')\GFkfs(\brd''',\brd'')d\brd'''d\brd'\,.
\end{multline}

In order to improve the numerical performance further I make a final few steps as in the original Born approximation \cite{Born26}, I define unit vector $\hat{\brd}$ such 
that $\brd=\rho\hat{\brd}$ and $\bk_s=k\hat{\brd}$. Then for $\rho>>\rho'$, 
\be
k|\brd-\brd'| = k\rho-\bk_s\cdot\brd'+{\it O}\left(\dfrac{1}{\rho^2}\right)+...
\label{ffa}
\ee
Therefore substituting \Eq{GF1} and \Eq{GFequ0fsSol} in to \Eq{eq:Mark0} and using \Eq{ffa} because both $\brd,\brd''$ are far from the scatterer I arrive at the RSE Born approximation
\begin{multline}\label{eq:Mark1}
\GFk(\brd,\brd'')=-\dfrac{i}{4}H_0(k|\brd-\brd''|)\hat{\mathbf{1}}
\\
-\omega^2\dfrac{Q^2e^{ik(\rho+\rho'')}}{\pi\sqrt{\rho\rho''}}\int e^{i(\bk_s-\bk_s'')\cdot\brd'}\Delta\heps_\omega(\brd')d\brd'
\\
+\omega^4\dfrac{Q^2e^{ik(\rho+\rho'')}}{\pi\sqrt{\rho\rho''}}\sum_n \frac{\An({\bk_s})\otimes\An(-{\bk_s}'')}{2k(k-k_n)}\,.
\end{multline}
or using \Eq{TUES_NIGHT1} instead
\begin{multline}\label{eq:Mark1GBA}
\GFk(\brd,\brd'')=-\dfrac{i}{4}H_0(k|\brd-\brd''|)\hat{\mathbf{1}}
\\
-\omega^2\dfrac{Q^2e^{ik(\rho+\rho'')}}{\pi\sqrt{\rho\rho''}}\int e^{i(\bk_s-\bk_s'')\cdot\brd'}\Delta\heps_\omega(\brd')d\brd'
\\
+\omega^4\dfrac{Q^2e^{ik(\rho+\rho'')}}{\pi\sqrt{\rho\rho''}}\sum_n \frac{\An({\bk_s})\otimes\An(-{\bk_s}'')}{2k_n(k-k_n)}\,.
\end{multline}

The vector $\An$ is defined as a Fourier transform of the RSs,
\be
\An(\bk_s)=\int e^{i\bk_s\cdot\brd'}\Delta\heps_\omega(\brd')\En(\brd')d\brd'
 \label{Ani}
\ee 
I note that the fast Fourier transform method is available. The first two terms in \Eq{eq:Mark1} correspond to the standard Born approximation, the final summation term corresponds to the 
RSE correction to the Born approximation.

A simple corollary of this theory is as follows, I can see from the arguments just stated that from Eq.(G5) if $\brd''$ 
is inside the resonator and $\rho>>\rho''$ then
\begin{multline}\label{eq:Mark3}
\GFk(\brd,\brd'')=-\dfrac{i}{4}H_0(k|\brd-\brd''|)\hat{\mathbf{1}}
\\
-\omega^2\dfrac{Qe^{ik\rho}}{\sqrt{\rho\pi}}\sum_n \frac{\An({\bk_s})\otimes\En({\brd}'')}{2k(k-k_n)}\,,
\end{multline}
or using \Eq{TUES_NIGHT1} instead
\begin{multline}\label{eq:Mark3GBA}
\GFk(\brd,\brd'')=-\dfrac{i}{4}H_0(k|\brd-\brd''|)\hat{\mathbf{1}}
\\
-\omega^2\dfrac{Qe^{ik\rho}}{\sqrt{\rho\pi}}\sum_n \frac{\An({\bk_s})\otimes\En({\brd}'')}{2k_n(k-k_n)}\,,
\end{multline}
similarly from Eq.(G6) if $\brd$ is inside the resonator and $\rho''>>\rho$ then
\begin{multline}\label{eq:Mark2}
\GFk(\brd,\brd'')=-\dfrac{i}{4}H_0(k|\brd-\brd''|)\hat{\mathbf{1}}
\\
-\omega^2\dfrac{Qe^{ik\rho''}}{\sqrt{\rho''\pi}}\sum_n \frac{\En({\brd})\otimes\An(-{\bk_s}'')}{2k(k-k_n)}\,,
\end{multline}
or using \Eq{TUES_NIGHT1} instead
\begin{multline}\label{eq:Mark2GBA}
\GFk(\brd,\brd'')=-\dfrac{i}{4}H_0(k|\brd-\brd''|)\hat{\mathbf{1}}
\\
-\omega^2\dfrac{Qe^{ik\rho''}}{\sqrt{\rho''\pi}}\sum_n \frac{\En({\brd})\otimes\An(-{\bk_s}'')}{2k_n(k-k_n)}\,,
\end{multline}
other permutations are possible.

It may be possible to use these formulas to develope a hollow glass cable with light and fluid passing along it, the escaping light being measured in the
far field for analysis with the RSE Born approximation for the determination of the content of the fluid via inverse emission problem. To elaborate on 
this point, it was shown in Ref.\cite{DoostARX15} that
\be
\label{GBA1}
\lim_{\rho\rightarrow\infty}\En(\brd)\propto\An(\hat{\brd}k_n)\dfrac{e^{ik_n\rho}}{\rho^{1/2}}\,.
\ee
However close to a sharp resonance, when [$k\approx\Re(k_n)$, $p$ is fixed], scattering is dominated by a single resonance and so the scattered ${\bf E}$-field ${\bf E}^{\rm scattered}$
is approximately 
\be
\label{GBA2}
\lim_{\rho\rightarrow\infty}{\bf E}^{\rm scattered}(\brd,k)\approx C\An(\hat{\brd}k)\dfrac{e^{ik\rho}}{\rho^{1/2}}\,.
\ee
Hence $\An(\hat{\brd}k)$ can be partial inverse Fourier transformed with respect to angle to find information about the internal structure of the cable, $C$ is some constant.
The same approach can be taken in three dimensions to aid the determination of resonator structure.

The reason \Eq{GBA1} takes the form it does can be seen more clearly by substituting the GF of \Eq{eq:Mark3} into the first line of \Eq{WHY} and taking the limit $k\rightarrow k_n$, while
following the arguments in that section that,
$$
\lim_{k\to {k_n}} \bE(\brd,k)=\En(\brd)\,.
$$

The 3D equivalent \cite{DoostARX15,Ge14} of \Eq{GBA1} one day might be valuable information for solving the inverse emission problem from such things as black hole gravitational wave emitters, 
i.e. calculations of the potential from the emission (decay) via fast inverse 
Fourier transform methods upon the set of $\An(\hat{\br}k_n)$, 
particularly if the potentials of interest are rotating about a fixed axis so we know their orientation to some extent such as occurs for decaying magnetic nuclei as part of a 
non-magnetic crystaline compound placed
inside a NMR (nuclear-magnetic-resonance) machine. Because $k_n$ are discrete values $\An(\hat{\br}k_n)$ only give angular information when inverse Fourier transformed
and so the inverse Fourier methods might have to be used self-consistently in conjuncture with
the RSE perturbation theory and the values of $k_n$. This is a highly speculative aside and might be a possible topic for future research.

\end{document}